\documentclass[aps,prb,twocolumn,groupedaddress]{revtex4-1}

\usepackage[utf8]{inputenc}
\usepackage{amsmath}
\usepackage{graphicx}
\usepackage{color}

\newcommand{\tmpnote}[1]%
   {\begingroup{\color{blue}\it (FIXME: #1)}\endgroup}

\newcommand\addtag{\refstepcounter{equation}\tag{\theequation}}
\DeclareRobustCommand{\orderof}{\ensuremath{\mathcal{O}}}

\begin{document}

\title{Intrinsic spin-orbit interaction in diffusive
  normal wire Josephson weak links: supercurrent
and density of states}

\author{Juho Arjoranta and Tero T. Heikkil\"a}
\affiliation{University of Jyvaskyla, Department of Physics and
  Nanoscience Center, P.O. Box 35, 40014 University of Jyv\"askyl\"a, FINLAND}

\date{\today}

\begin{abstract}
We study the effect of the intrinsic (Rashba or Dresselhaus) spin-orbit interaction in
superconductor--nanowire--superconductor (SNS) weak links in the presence of a spin-splitting
field that can result either from an intrinsic exchange field or the
Zeeman effect of an applied field. We solve the full non-linear
Usadel equations numerically \cite{numericsnote} and analyze the resulting supercurrent
through the weak link and the behavior of the density of states in the
center of the wire. We point out how the presence of the
spin-orbit interaction gives rise to a long-range
spin triplet supercurrent, which remains finite even in the limit of
very large exchange fields. In particular, we show how rotating the
field leads to a sequence of transitions between the 0 and $\pi$
states as a function of the angle between the exchange field and the
spin-orbit field. Simultaneously, the triplet
pairing leads to a zero-energy peak in the density of states. We proceed by solving the linearized
Usadel equations, showing the correspondence to the solutions of the
full equations and detail the emergence of the long-range supercurrent
components. Our studies are relevant for on-going investigations of
supercurrent in semiconductor nanowires in the limit of several
channels and in the presence of disorder.
\end{abstract}


\maketitle

\section{Introduction}
The antagonist nature of conventional singlet superconductivity and magnetism has been
clearly illustrated in experiments studying supercurrents
flowing through ferromagnetic weak links \cite{ryazanov01,kontos01}. There, the spin-splitting
(exchange) field $h$ suppresses the supercurrent within a typically short
magnetic length scale $\xi_m=\sqrt{\hbar D/h}$, where $D$ is the
diffusion constant of the wire. As suggested in
Ref.~\onlinecite{bergeret01}, this suppression can be lifted by
converting part of the singlet supercurrent into a triplet with a
finite projection of the magnetic moment of Cooper pairs, by utilizing
an inhomogeneous magnetization at the
interface between the ferromagnet and the superconductor. This
component couples electrons with spins from the same band, and
therefore it is not sensitive to the spin-splitting field. This
suggestion was experimentally demonstrated in a number of works
\cite{khaire10,robinson10} utilizing a series of different types of
magnetic layers that are non-collinear with respect to each other.

\begin{figure}[h]
\includegraphics[width=\columnwidth]{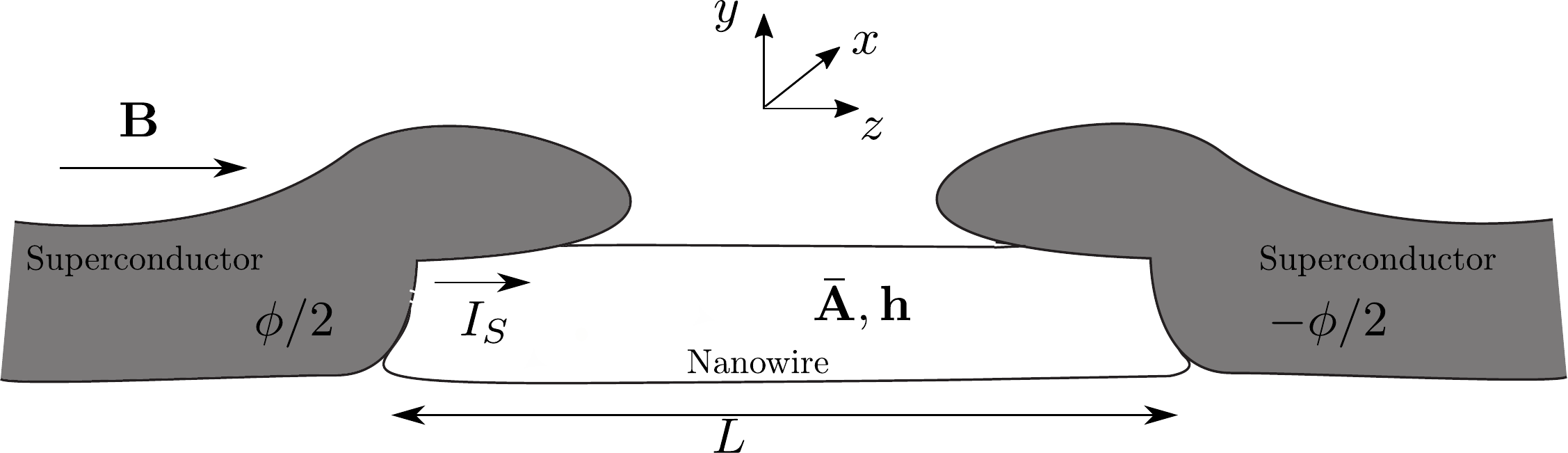}
\caption{SNS junction studied in this work: a diffusive nanowire
  of length $L$ connects two bulky superconductors. The nanowire is
  characterized by the spin-orbit field $\mathbf{\bar A}$ and the
  system exhibits an exchange field $\mathbf h$ either due to an applied
  magnetic field $\vec{B}$ or by proximity to magnetic material (not
  in the picture). We assume that both $\mathbf{\bar A}$ and
  $\mathbf{h}$ are in the $xz$-plane, and that $\mathbf{h}$ is at an angle
  $\theta$ compared to the ($z$) direction of the wire. \label{fig:SNS-junction}}
\end{figure}

Besides using magnetic materials, the spin-splitting field can be
realized via the Zeeman effect of an applied magnetic field.\cite{heikkila00,yip00} Similar
physics as in the SFS case can be envisaged as long as the orbital
effect of the magnetic field is weak enough \cite{crosser08,cuevas07} and does
not limit the supercurrent. Such a situation takes place especially in
narrow nanowires, where the spin-splitting field in combination with
the Rashba-type spin-orbit (SO) interaction has been used in an effort to
take these wires to the limit of topological superconductivity
\cite{Science.336.1003,das12,churchill13} for the detection of Majorana-type excitations at
the edges of the wires. Most of such experiments are nevertheless in
the topologically trivial limit. It is hence of interest to study the
physics of such nanowires in the presence of the combination of the
spin-orbit and spin-splitting fields. This is the aim of the present
work. In particular, we study the supercurrent behavior in systems
schematically presented in Fig.~\ref{fig:SNS-junction}. Contrary to
many recent theory works on the effects of spin-orbit coupling on
proximity superconductivity discussing the fully ballistic
regime,\cite{PhysRevLett.104.040502,oreg2010,sun2015} we assume the
wires to be diffusive. Strictly speaking this limit requires that all
wire dimensions are smaller than the elastic mean free path. Typical
epitaxial nanowires have mean free paths comparable to the wire
thickness, and much less than the wire
length.\cite{privatecommunication,paajaste2015,giazotto2011,spathis2011,roddaro2011,vanweperen2013,gul2015,vanweperen2015}
Even in this limit the diffusive-limit theory is likely to capture the
essential physics much better than the fully ballistic limit. In the
diffusive limit it is generally possible to obtain a fully
quantitative fit with between theory and experiments,\cite{dubos2001,lesueur2008} which is why
also the quantitative details of the theory are relevant. On the
other hand, the quasiclassical theory we employ corresponds to setting
the Fermi wavelength $\lambda_F \rightarrow
0$. Therefore, it cannot capture effects related to for example weak
antilocalization, possibly relevant in these wires.\cite{bagrets2012,pikulin2012,vanweperen2015}
Alternative derivations of the quasiclassical theory on the fixed
few-channel limit \cite{neven2013} cannot be directly connected on the
many-channel limit considered here.

This paper extends on the work of Tokatly and Bergeret,\cite{PhysRevLett.110.117003, PhysRevB.89.134517}
who introduced the mechanism of including the intrinsic (Rashba or
Dresselhaus) spin-orbit interaction as a spin-dependent vector
potential into the Usadel equation describing inhomogeneous
superconductivity in the diffusive limit. They also pointed out how
for certain relative orientations of the wire, spin-orbit fields and
the exchange field, the combination of the latter two may produce
triplet supercurrent that survives even at large exchange fields. In
particular, they showed that the wire with a homogeneous exchange
field and intrinsic spin-orbit interaction is gauge equivalent to a
ferromagnet with inhomogeneous magnetization (see also Ref.~\onlinecite{gorini10}). Here
we study this mechanism quantitatively (Sec.~\ref{sec:supercurrent}). In particular, we show the
dependence of the supercurrent vs. exchange field for varying
magnitudes of the Rashba field. We also demonstrate in detail how the
triplet supercurrent depends on the direction of the magnetic field
applied in the plane of the wire. For strong spin-orbit coupling,
we predict that changing the direction of the field drives the junction through a
sequence of 0-$\pi$ transitions.

Besides supercurrent, we also study the local density of states in the
junction in Sec.~\ref{sec:dos}. This is also the typical observable in the studies of
Majorana physics. We find that spin-orbit interaction induces a
zero-energy peak for a range of exchange fields. This peak originates
from the induced long-range triplet amplitude of the superconducting
pairing, and it is quite
sensitive to the precise direction of the field and the amplitude of
the SO coupling.

\section{Usadel equation with spin-orbit coupling}

We implement the spin-orbit interaction into the Usadel
equation describing the quasiclassical Nambu-spin retarded Green's
function $\hat G^R$ in the diffusive limit\cite{Rev.Lett.25.507,
  SuperlatticesMicrost.25.1251,PhysRevLett.110.117003} (here and
below, $e=\hbar=k_B=1$ except when we discuss particular values of the
observables) 
\begin{equation}
D\hat{\nabla}\cdot\left(\hat{G}^R\hat{\nabla}\hat{G}^R\right)=\left[-i\varepsilon\hat{\tau}_{3}-i\mathbf{h}
  \cdot \mathbf{\bar\sigma}+\hat{\Delta}+\hat{\Sigma}_{\rm
      sf}+\hat{\Sigma}_{\rm in},\,\hat{G}^R\right]. \label{eq:usadel1}
\end{equation}
Here $D$ is the diffusion constant inside the nanowire, $\varepsilon$
is the energy, $\mathbf{\bar \sigma}$ is a vector of Pauli matrices in spin
space, ${\mathbf h}=(h_x,h_y \hat \tau_3,h_z)$ and  
$\hat{\nabla}\hat{G}^R=\nabla_{\mathbf{R}}\hat{G}^R-i\left[\mathbf{\bar{\textbf{A}}}\hat{\tau}_{3},\,\hat{G}^R\right]$
is the gauge invariant gradient with $\bar{\mathbf{A}}$ describing the SO
coupling. The latter is specified below in more detail.\cite{representationnote}

Inside the normal metal, the superconducting pair potential
$\hat{\Delta}$ is zero and we assume that the term
$\hat{\Sigma}_{\rm in}$ describing the inelastic scattering is
negligible. In practice, the junction contains also regular spin-flip
scattering and scattering due to isotropic spin-orbit coupling,\cite{abrikosov62}
characterized by the self-energy $\hat \Sigma_{\rm sf}$. We assume
spin relaxation to be dominated by the intrinsic SO coupling and
neglect these other terms in the following. Thus inside the normal metal we have 
 \begin{equation}
D \hat{\nabla} \cdot ( \hat{G}^R \hat{\nabla} \hat{G}^R ) = [-i
\varepsilon^+ \hat{\tau}_3 - i\mathbf{h} \cdot \mathbf{\bar{\sigma}}, \hat{G}^R ],
\label{eq:usadel2}
\end{equation}
where $\varepsilon^+=\varepsilon+i\eta$, and $\eta \rightarrow 0^+$ is
a small term specifying the location of the poles of the Retarded
Green's function.

Introducing a dimensionless position coordinate $z' = z / L$ and
defining a Thouless energy $E_T = D / L^2$, where $L$ is the length of
the normal metal wire, we can work in energy units of $\varepsilon = E
/ E_T$ and $\mathbf{\tilde h}=\mathbf h/E_T$ and use the scaled vector
potential $\bar{\textbf{A}}_s = \bar{\textbf{A}}L$. Moreover, below we
assume the Zeeman field to lie in the substrate plane, and therefore $h_y=0$. Equations \eqref{eq:usadel1} and \eqref{eq:usadel2} have to be
supplemented with the normalization condition $(\hat G^R)^2=\hat 1$. In the numerical solutions,
we implement this by using the Riccati parameterization
\cite{Phys.Rev.B.61.9061} (see Appendix A) for the retarded Green's
function. In that parametrization, the Nambu-space Green's function is
specified in terms of two parameters, $\gamma$ and $\tilde \gamma$. In general, both of these parameters
are $2 \times 2$ matrices in spin space. The Usadel equation written
for $\gamma$ reads
\begin{multline}
\partial_{z'}^{2} \gamma - 2 (\partial_{z'} \gamma) \tilde{\gamma} N (\partial_{z'} \gamma) = \\
-2 i \epsilon \gamma - i\mathbf{\tilde h} \cdot [\gamma, \mathbf{\bar{\sigma}}] + [\bar{\textbf{A}}_s^2, \gamma] + 2 \{\bar{\textbf{A}}_s, \gamma\} \tilde{N} (\bar{\textbf{A}}_s - \tilde{\gamma} \bar{\textbf{A}}_s \gamma) \\
+ 2i \left((\partial_{z'} \gamma) \tilde{N} (\bar{\textbf{A}}_s - \tilde{\gamma} \bar{\textbf{A}}_s \gamma) + (\bar{\textbf{A}}_s - \gamma \bar{\textbf{A}}_s \tilde{\gamma}) N (\partial_{z'} \gamma) \right). \label{UsadelRiccati}
\end{multline}
For equilibrium observables, we replace $-i \epsilon$ by the Matsubara
frequencies $\omega_n = \pi T (2n + 1)$.\cite{Matsubara} The equation
for $\tilde{\gamma}$ is obtained by substituting $\gamma
\leftrightarrow \tilde{\gamma}$, $N \leftrightarrow \tilde{N}$ and by
taking a complex conjugate of the scalars.

\subsection{Spin-orbit field}
A generic spin-orbit coupling emerging in systems with broken
inversion symmetry is of the form $\bar{A}_i=\alpha_i^j \bar{\sigma}_j$. For a
one-dimensional wire the symmetry is broken by the geometry in both
directions perpendicular to the wire, and therefore the components
$\alpha_i^j$ can be considered independent of each other. 
To describe a one-dimensional wire in the $z$ direction placed on a substrate (see
Fig.~\ref{fig:SNS-junction}) spanning the $xz$ plane, we describe the
generic spin-orbit field as 
\begin{equation}
\bar{\mathbf{A}}=\begin{pmatrix} \alpha_1^1 \bar{\sigma}_1 + \alpha_1^3 \bar{\sigma}_3, & 0, &
  \alpha_3^1 \bar{\sigma}_1 + \alpha_3^3 \bar{\sigma}_3 \end{pmatrix}.
\end{equation}
For a thin wire there are gradients only in the $z$ direction, and
therefore the only terms coupling linearly to the gradient are related
to $\alpha_3$. The other terms can be non-vanishing and they contribute in general
to the Dyakonov-Perel spin relaxation \cite{PhysRevLett.110.117003, PhysRevB.89.134517},
but as they do not lead to other interesting physics, we neglect them
in our numerical results. After neglecting the orbital effect of the
field, the properties of the junction depend only on the relative
direction between $\mathbf{h}$ and $\vec{A}_z$ in $\bar A_z =
\vec{A}_z \cdot \vec\sigma$. As we vary the direction of $\mathbf h$,
we fix $\bar A_z \propto \sigma_1$, i.e., $\alpha_3^3=0$. We vary the remaining term $\alpha=\alpha_3^1$ to
observe the physics due to the intrinsic spin-orbit coupling. Note
that the term $\alpha \sigma_1 \hat u_z$ gives rise to the linear
energy term of the form $e\alpha \sigma_1 p_z/m$. Our term $\alpha$ is
related to the 
(Rashba) terms of size $\tilde \alpha$ considered in the literature on Majorana fermions (e.g.,
\cite{cayao15}) by
\begin{equation}
\alpha = \frac{m\tilde \alpha}{\hbar e},
\end{equation}
where $m$ is the effective mass of the electrons inside the
nanowire. Our $\alpha$ hence has a dimension of $\hbar/(e L)$, where
$L$ is some length scale. In the numerics, the important dimensionless
parameter is $e\alpha L/\hbar$, where $L$ is chosen to be the distance
between the two nanowire-superconductor contacts. For example with the parameters of InSb discussed in
Ref.~\onlinecite{cayao15} ($m=0.015 m_e$ and $\tilde \alpha=0.2$ eV
\AA, where $m_e$ is the electron mass), we get $e \alpha/\hbar
\approx 4 \cdot 10^6$ 1/m. These estimates are consistent with the
recent experiments \cite{vanweperen2015}, which obtain even a
somewhat larger value of $\tilde \alpha$. For typical wires of length $L \sim $ 1
$\mu$m, the value of the dimensionless parameter can hence be of the
order of or larger than unity. On the other hand, in order to be able
to neglect gradients in transverse directions in 
Eq.~\eqref{UsadelRiccati}, we assume that the wires are narrow
compared to the spin-orbit length, i.e., wire thickness $d$ satisfies
$d \ll \hbar/e\alpha$. 

\subsection{Boundary conditions}

For the boundary conditions to the Green's functions, we assume clean NS interfaces. This assumption means that the parameters $\gamma$ and $\tilde{\gamma}$ are continuous across the interface and coincide with bulk BCS superconductor values \cite{SuperlatticesMicrost.25.1251, RevModPhys.58.323}

\begin{subequations}
\begin{align}
\gamma_{BCS}=\frac{i\left|\Delta\right|e^{i\varphi}}{\varepsilon+i\sqrt{\left|\Delta\right|^{2}-\left(\varepsilon+i0^{+}\right)^{2}}}\bar\sigma_2\\
\tilde{\gamma}_{BCS}=\frac{i\left|\Delta\right|e^{-i\varphi}}{\varepsilon+i\sqrt{\left|\Delta\right|^{2}-\left(\varepsilon+i0^{+}\right)^{2}}}\bar\sigma_2
\end{align}
\label{Keldysh_boundary}
\end{subequations}
in the real-time description (for Matsubara frequencies, replace
$\epsilon$ by $i\omega_n$). The exact form of the boundary conditions
depends on the chosen form of the Nambu vector. 

In the numerics, we express all the lengths in terms of the length $L$
of the wire. In this case, the natural energy scale is given by the
Thouless energy $E_T=\hbar D/L^2$. We mostly concentrate on the limit
of long wires, where $L \gg \xi_0=\sqrt{\hbar D/\Delta}$. This also
means that $\Delta/E_T = L^2/\xi_0^2 \gg 1$.

\subsection{Supercurrent}
The supercurrent through the junction is characterized by the spectral current density \cite{Heikkila02}

\begin{equation}
 j_{s}=\frac{L}{4}\text{Tr}\left[\left(\hat{G}^{R}\hat{\nabla}\hat{G}^{R}-\hat{G}^{A}\hat{\nabla}\hat{G}^{A}\right)\hat{\tau}_{3}\right].
\end{equation}

\noindent In the Riccati parameterization, the spectral current
density can be written as 

\begin{align*}
j_{s}=\frac{L}{2}&\left\{ \text{Tr}\left[ N\left(\gamma\tilde{\gamma}'-\gamma'\tilde{\gamma}\right)N-\tilde{N}\left(\tilde{\gamma}\gamma'-\tilde{\gamma}'\gamma\right)\tilde{N}\right.\right.\\
&\left.\left.+i\bigg(N\left\{ \bar{\textbf{A}},\gamma\right\} \tilde{\gamma}N+N\gamma\left\{ \bar{\textbf{A}},\tilde{\gamma}\right\} N \right.\right.\\
&\left.\left.+\tilde{N}\left\{ \bar{\textbf{A}},\tilde{\gamma}\right\} \gamma\tilde{N}+\tilde{N}\tilde{\gamma}\left\{ \bar{\textbf{A}},\gamma\right\} \tilde{N}\bigg)\right]\right\}, \addtag \label{spectral_current_density_riccati}
\end{align*}
where $\gamma'=\partial \gamma/\partial z$. 

The supercurrent can be calculated as a weighted average of the
spectral current density

\begin{equation}
I_{s}=\frac{E_T}{2 e R_N}\int_{-\infty}^{\infty}d\varepsilon
{\rm Re} j_{s}\left(\varepsilon\right) \text{tanh}\left(\frac{\varepsilon}{2T}\right),
\label{eq:observablesupercurrent1}
\end{equation}

\noindent where $R_N=L/(A e^2 D \nu_F)$ is the Drude resistance of the
nanowire in the normal state. Here $L$ is the length, $A$ is the
area, $D$ the diffusion constant and $\nu_F$ the density
of states at the Fermi level of the nanowire in its normal state. In
the Matsubara technique, the integral can be calculated as a sum of
the spectral current densities evaluated at the Matsubara frequencies
$\omega_n$ as

\begin{equation}
I_{s}=-\frac{E_T}{e R_N}2\pi
T\sum_{n=0}^{\infty}{\rm Im} j_{s}\left(\omega_{n}\right).
\label{eq:observablesupercurrent}
\end{equation}

In the numerics, we cut the Matsubara sum to the index after which the
obtained supercurrent changes less than 1 \%.

\subsection{Density of states}
The density of states (DOS) is given by \cite{SuperlatticesMicrost.25.1251}
\begin{equation}
N(\varepsilon, \mathbf{R}) = \frac{N_F}{2} \text{Re} \Bigg\{ \text{Tr} \bigg[ \hat{G}^R(\varepsilon, \mathbf{R}) \hat{\tau}_3 \bigg] \Bigg\},
\end{equation}
\noindent where $N_F$ is the DOS in the absence of
superconductivity. In the Riccati parameterization,
\begin{equation}
N\left(\varepsilon,\mathbf{\,
    R}\right)=\frac{N_{F}}{2}\text{Re}\text{Tr}\left\{
  N\left(1-\gamma\tilde{\gamma}\right)+\tilde{N}\left(1-\tilde{\gamma}\gamma\right)\right\}.
\label{eq:ldos}
\end{equation}
Besides supercurrent, the density of states is another way to characterize the excitation
spectrum induced by the proximity effect from the supercurrent. It can
be accessed via a standard tunneling measurement. In
Sec.~\ref{sec:dos} we show how a combination of the finite exchange
field and the Rashba spin-orbit coupling gives rise to a zero-energy
peak in the density of states. This is qualitatively similar to what
one expects from the measurements in the Majorana wires,\cite{Science.336.1003,das12,churchill13} although the
physics of this effect is quite different. Initial results for the
density of states in Rashba wires were presented by us in
\onlinecite{juhongradu}. Recently, similar type of results were
discussed also in Refs.~\onlinecite{jacobsen15a,jacobsen15b}, but
that work concentrated on superconductor/ferromagnet multilayers in
the short junction limit, where the junction length is of the order of
the superconducting coherence length $\xi_0=\sqrt{\hbar D/\Delta}$. In
such multilayers, the emergence of the long-range triplet
superconductivity requires either the presence of both Rashba and
Dresselhaus type spin-orbit coupling, or out-of-plane magnetic
fields. Moreover, for short junctions the energy scales are
primarily set by the superconducting gap $\Delta$ instead of the Thouless
energy as here. Nevertheless, also there the triplet proximity effect
leads to the presence of zero-energy density of states peaks, and
long-range supercurrent. However, the quantitative details of the
results are quite different, and therefore as such not applicable for
the nanowire setups.

\section{Generation of the long-range triplet component}

Below, we present numerical solutions of the supercurrent and density
of states in a nanowire Josephson weak link exhibiting both
intrinsic spin-orbit coupling and exchange field. To understand
these results, let us first study what we expect to find in the limit
of a weak proximity effect, \cite{PhysRevLett.110.117003,
  PhysRevB.89.134517} which allows us to linearize the Usadel equation.

Linearizing Eq.~\eqref{UsadelRiccati} yields 

\begin{equation}
\partial_{z'}^{2} \gamma = 2 \omega_n \gamma - i\mathbf{\tilde h} \cdot
[\gamma, \mathbf{\bar{\sigma}}] + [\bar{\textbf{A}}_s^2, \gamma] + 2 \{\bar{\textbf{A}}_s, \gamma\}{\bar{\textbf{A}}_s} + 2i \{\partial_{z'} \gamma, \bar{\textbf{A}}_s\}. \label{linearized_Usadel}
\end{equation}

\noindent Here we have discarded all the terms \orderof($\gamma^2$)
and noticed that $N = 1$ due to the normalization condition
$\left(\hat{G}^{R}\right)^{2}=1$. We separate the different spin
components as
\begin{equation}
\gamma = \sum_{i=0}^{3} f_i \bar{\sigma}_i
\end{equation}
\noindent and choose a specific SOC field $\bar{\textbf{A}}_s = \alpha \bar{\sigma}_1 \hat
u_z$ and direction of the magnetic field, $\mathbf{\tilde
  h}=h(0,\sin(\theta),\cos(\theta))$. The resulting equations describe the interplay between
the short-range and long-range pairing components. 

We can solve the resulting second-order boundary value problem by
separating the short-range and long-range components $f_i$, and using
the previous to find effective boundary conditions for the latter. The details are given in 
Appendix B. As a result, we get the spectral supercurrent describing
the long-range component. In the case of a perpendicular field and
$\bar{\mathbf A}_s$ ($\theta=0$), the result reduces to
\begin{equation}
{\rm Im} (j_s) =-\frac{|\Delta|^2}{(\omega_n+\sqrt{\omega_n^2+|\Delta|^2})^2}\frac{8\sqrt{2}
  \alpha^2 D \sqrt{\tilde{\omega}_n}}{h
  \sinh(\sqrt{2}\sqrt{\tilde{\omega}_n})}\sin(\phi),
\label{eq:spectralscanalytic}
\end{equation}
where
$\phi$ is the phase difference between the two
superconductors, and $\tilde \omega_n=(\omega_n+2D\alpha^2)/E_T$ is
the (dimensionless) Matsubara frequency modified by the pair-breaking effect from the
spin-orbit coupling ($\omega_n\mapsto -i\epsilon$ in the real-time
formulation with energy $\epsilon$). Note that the spin-orbit coupling
$\alpha$ plays here a dual role: first, it induces the long-range
triplet component of the (spectral) supercurrent, and second, it
induces pair breaking effects. If we would include the ordinary
spin-flip or spin-orbit scattering effects, they would (in
the lowest order) induce the corresponding terms into $\tilde \omega$.  

The observable supercurrent at a given temperature $T$ is obtained by
summing over the Matsubara frequencies or integrated over the real
energies as in
Eqs.~(\ref{eq:observablesupercurrent1},\ref{eq:observablesupercurrent}). At
low temperatures $k_B T \lesssim E_T$ we may transform the Matsubara
sum into an integral and obtain in the long-junction limit $E_T, e^2 \hbar \alpha^2 D \ll
\Delta$ (restoring $e$ and $\hbar$)
\begin{equation}
\label{eq:supercurrentanal}
\begin{split}
&I_S(T \approx 0)=\\&\frac{32 E_T e \alpha^2 D}{hR_N \hbar} e^{-2 \alpha
  L e/\hbar} \left(1+\frac{2 \alpha L e}{\hbar}+\frac{2 \alpha^2 L^2
    e^2}{\hbar^2}\right) \sin(\phi).
\end{split}
\end{equation}
Strictly speaking this is valid only for $2e\alpha L/\hbar \gtrsim 1$,
but this approximation fits numerical results reasonably well also for
low $\alpha$. On the other hand, at high temperatures $k_B T \gg E_T$,
it is enough to include only the lowest Matsubara frequency and the
result is
\begin{equation}
\begin{split}
&I_S(T \gg E_T)=\\&\frac{|\Delta|^2}{(\pi k_B T + \sqrt{\pi^2
    k_B^2T^2+|\Delta|^2})^2}\frac{8  E_T e \alpha^2 D}{h R_N \hbar}
\sqrt{x} e^{-\sqrt{x}},
\end{split}
\end{equation}
where $x=2 \pi k_B T/E_T + 2 \alpha L e/\hbar$. 

We compare the
analytical result to the full numerics in
Fig.~\ref{fig:alpha_dependence}. For large $\alpha \gtrsim 6 \hbar/(eL)$, the full theory
shows a second maximum in the supercurrent, absent in this analytical
approximation (see Fig.~\ref{fig:alpha_dependence}).

For a non-zero $\theta \in ]0,\pi/2[$, both long-range components
$f_0$ and $f_3$ become non-zero. Due to their coupling, the
supercurrent obtains terms oscillating with $\alpha Le\sin(\theta)/\hbar$. The resulting oscillations vs. $\theta$
for $\alpha L e \gtrsim 1$ signal transitions between $0$ and $\pi$ states, which would be
observed as cusps in the dependence of the critical current on the
field direction. This is qualitatively described in Appendix B and further explored in
Fig.~\ref{fig:supercurrent_exchange_field_angular_dependence}. 

\section{Supercurrent}
\label{sec:supercurrent}
The effects of the spin-orbit coupling show up only in the presence of
a non-zero exchange field in the wire. This exchange field can be
established either in an intrinsically ferromagnetic wire, via the
magnetic proximity effect in a wire in contact with a ferromagnetic
insulator, or via the Zeeman field from an applied external magnetic field. In
all these cases the (direct or stray) magnetic field generates also
the orbital effect for the charge carriers. The relevance of this
orbital effect depends on the aspect ratio of the wire
\cite{crosser08,cuevas07}. Here we assume that the wire is thin enough so
that we can disregard the orbital effect and concentrate only on the
Zeeman field. Unless stated otherwise, the numerical results are obtained for a field in
the direction of the wire (i.e., $\theta=0$).

\begin{figure}[h!]
\includegraphics[width=0.9\columnwidth]{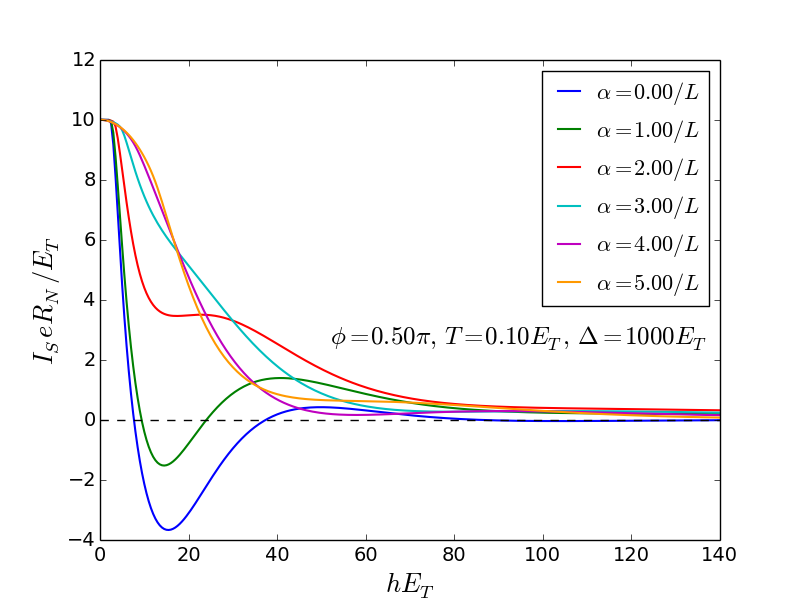}
\caption{Exchange field dependence of the supercurrent in a SNS
  junction for different intrinsic spin-orbit coupling strengths $\alpha$. A finite SO
  coupling strength yields a finite supercurrent through the junction
  even with large exchange fields. \label{exchange_dependence_different_alpha}}
\end{figure}

The most straightforward experiment is to vary the magnitude of the
exchange field. In Fig.~\ref{exchange_dependence_different_alpha} we
plot the supercurrent $I_S$ at phase difference $\phi=\pi/2$ (close to the
critical current) at the temperature $T=0.1 E_T$ for a long junction
(length $\Delta/E_T=L^2/\xi_0^2=1000$) as a function of the exchange field for varying
magnitudes of the spin-orbit coupling $\alpha$. For low $\alpha
\lesssim 1/L$, the supercurrent exhibits damped oscillations as a
function of the exchange field as shown before for example in
Refs.~\onlinecite{buzdin82,ryazanov01}. For $I_S <0$, the junction enters the
$\pi$ state. With an increasing spin-orbit field, the oscillation of the
supercurrent is supressed and the junction no longer can be found in
the $\pi$-state (note that non-quasiclassical corrections to our
theory can lead to the presence of a $\phi$-state\cite{krive04,bergeret14} with $\phi \neq
0,\pi$ due to the spin-orbit coupling). In the absence of spin-orbit
coupling, $\alpha=0$, the supercurrent dies out for large exchange
fields. However, a finite $\alpha$ yields also a finite supercurrent
even at rather large exchange fields. 

\begin{figure}[h!]
\includegraphics[width=0.9\columnwidth]{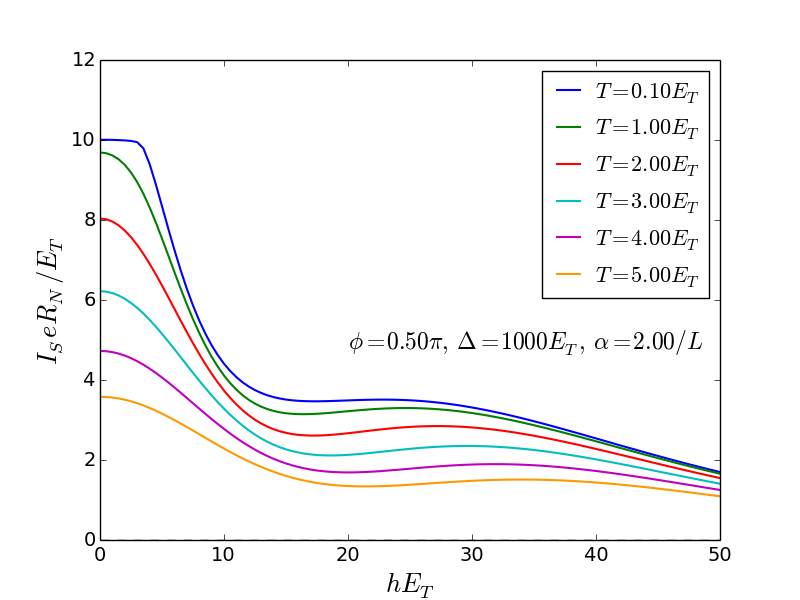}
\caption{Supercurrent through the SNS junction as a function of the exchange field with different temperatures. \label{fig:exchange_dependence_different_T}}
\end{figure}

For an intermediate value of $\alpha=2/L$, the supercurrent
vs.~exchange field exhibits a minimum at $h \approx 10 \dots 20
E_T$. This minimum persists to rather high temperatures as shown in
Fig.~\ref{fig:exchange_dependence_different_T}. 

\begin{figure}[h!]
\includegraphics[width=0.9\columnwidth]{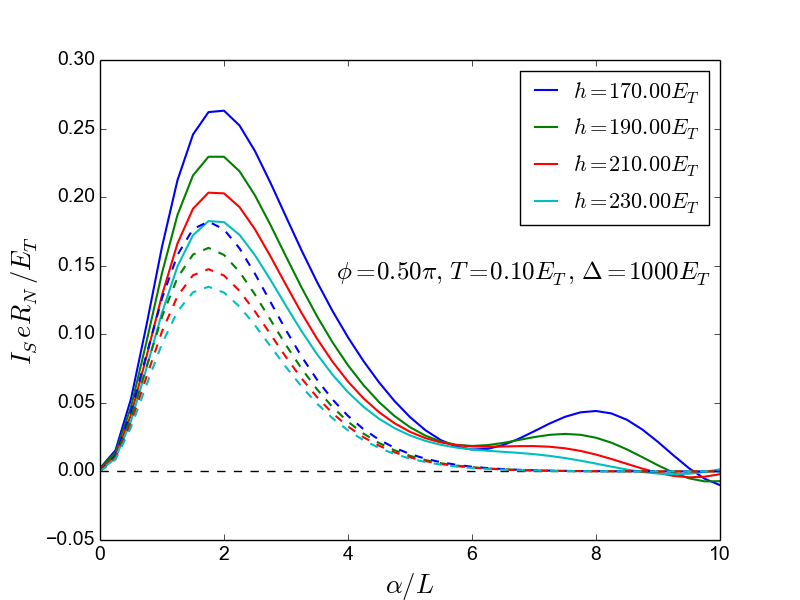}
\caption{With large exchange fields, the supercurrent oscillates as a
  function of the spin-orbit field strength reaching a maximum around
  $\alpha = 2 / L$. Solid lines show the numerical results and dashed
  lines are obtained by using Eq.~\eqref{eq:spectralscanalytic} in Eq.~\eqref{eq:observablesupercurrent}. \label{fig:alpha_dependence} }
\end{figure}

We can compare our results quantitatively to those obtained from the
linearized Usadel equation at large values of the exchange field,
where the major contribution to the supercurrent comes from the
long-range components. In Fig.~\ref{fig:alpha_dependence} we plot the
supercurrent for a few values of the exchange field as a function of
$\alpha$. The solid lines show the results from the exact numerics,
whereas the dashed lines come from the Matsubara
sum of Eq.~\eqref{eq:spectralscanalytic} (quite close to Eq.~\eqref{eq:supercurrentanal}). We find that both approaches yield a supercurrent
that is non-monotonous with respect to the value of the spin-orbit
field. Equations
(\ref{eq:spectralscanalytic},\ref{eq:supercurrentanal}) capture
the first oscillation rather well, but the amplitude of the exact
supercurrent is somewhat larger that obtained from the analytics.

\begin{figure}[h!]
\includegraphics[width=0.9\columnwidth]{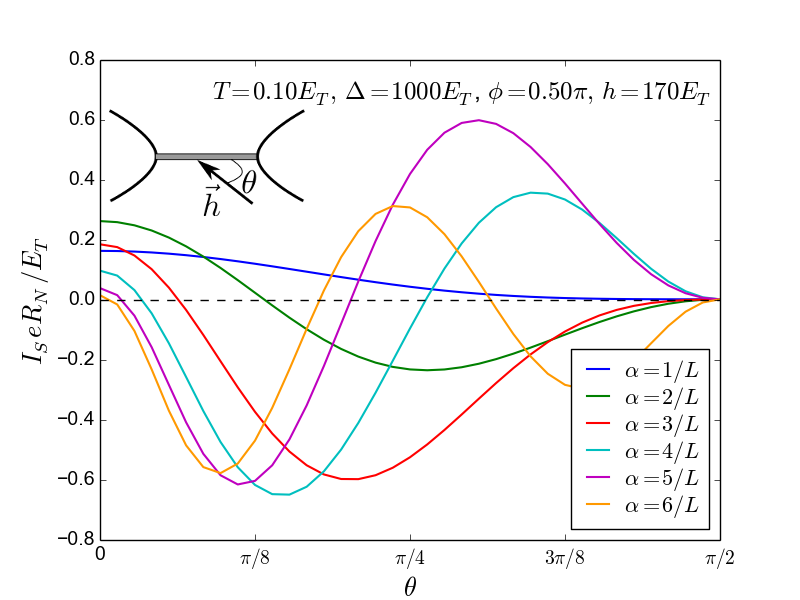}
\caption{Supercurrent as a function of the angle $\theta$ of the applied field,
  $\bar{h} = h (\text{cos}\theta \bar\sigma_3 +
  \text{sin}\theta \bar\sigma_1)$, $h=170E_T$ and otherwise the same
  parameters as in Fig.~\ref{exchange_dependence_different_alpha}. Rotating the exchange field makes the
  long-range component disappear when the spin-orbit field is parallel
  to the exchange field ($\theta=\pi/2$). In addition, applying a
  field at an intermediate angle results into an alternating sequence
  of $0-\pi$ transitions, their number depending on the precise value
  of $\alpha$. \label{fig:supercurrent_exchange_field_angular_dependence} }
\end{figure}

Tokatly and Bergeret \cite{PhysRevLett.110.117003, PhysRevB.89.134517} showed that in order to get a long-range triplet
supercurrent, the term describing the exchange field should not
commute with the vector potential. We check this by investigating the
dependence of the supercurrent on the angle $\theta$ of an in-plane exchange field with
respect to the wire direction (see inset of Fig.~\ref{fig:supercurrent_exchange_field_angular_dependence}). For $\theta=0$ the exchange field is in the $z$-direction and
therefore produces the term proportional to $\bar\sigma_3$, whereas we
describe the (Rashba-type) spin-orbit term proportional to $\bar\sigma_1$. For
$\theta=\pi/2$, both are described by terms proportional to
$\bar\sigma_1$. The supercurrent $I_S(\theta)$ is plotted in
Fig.~\ref{fig:supercurrent_exchange_field_angular_dependence}. Indeed,
for $\theta=\pi/2$, $I_S(\theta)$ only contains singlet components of
the pairing amplitude and therefore almost vanishes because of the
large value of the chosen exchange field. In addition, as described in
Appendix B, we find that for $\alpha \gtrsim 1/L$, the junction shows
a sequence of $0-\pi$ transitions, signalled by the sign change of
supercurrent at $\phi=\pi/2$. The magnitude of $\alpha$ dictates the
position of these transitions. For large exchange field, these
positions do not depend at all on the field, and they only weakly
depend on temperature. In experiments, $\alpha$ is not usually
a variable quantity. However, the angle of the applied field can be
varied straightforwardly. Therefore, studying the detailed angular
dependence of the supercurrent, one is able to determine the magnitude
and direction of the Rashba vector potential $\mathbf{\bar A}_s$. 

\section{Density of states}
\label{sec:dos}
Besides supercurrent, the proximity effect from the superconductor on
the normal wire can be characterized via a tunneling measurement of
the density of states. In the absence of the exchange field or the
spin-orbit field, the density of states in the proximity wire exhibits
a phase-dependent minigap,\cite{zhou98} whose size can be
approximatively described via $E_g(\phi) \approx 3.1 E_T
\cos^2(\phi/2)$. In the absence of spin mixing (either via the
intrinsic spin-orbit coupling considered here, spin-flip scattering or the isotropic
spin-orbit coupling), the exchange field simply shifts this minigap
by $\sigma h$ for spin $\sigma=\pm$. For $h > E_g(\phi)$, we hence
expect to see two regions with $N(E)=1/2$ in the spin-averaged density
of states $N(E)$. The intrinsic spin-orbit coupling mixes the
spins and leads to a closing of the minigaps for the spin-resolved
density of states. As we show below, an intermediate magnitude of the intrinsic spin-orbit
coupling also leads to a zero-energy peak in the density of
states. The shape and height of this peak is very sensitive to the
exact parameters of the system. We illustrate these results in the
following via a few examples, a more complete description can be found
from Ref.~\onlinecite{supplement}. 

The density of states is in general position dependent. To illustrate
the proximity induced effects in a symmetry point, we present the
results calculated in the middle of the normal metal.

\begin{figure}[h]
\includegraphics[width=0.9\columnwidth]{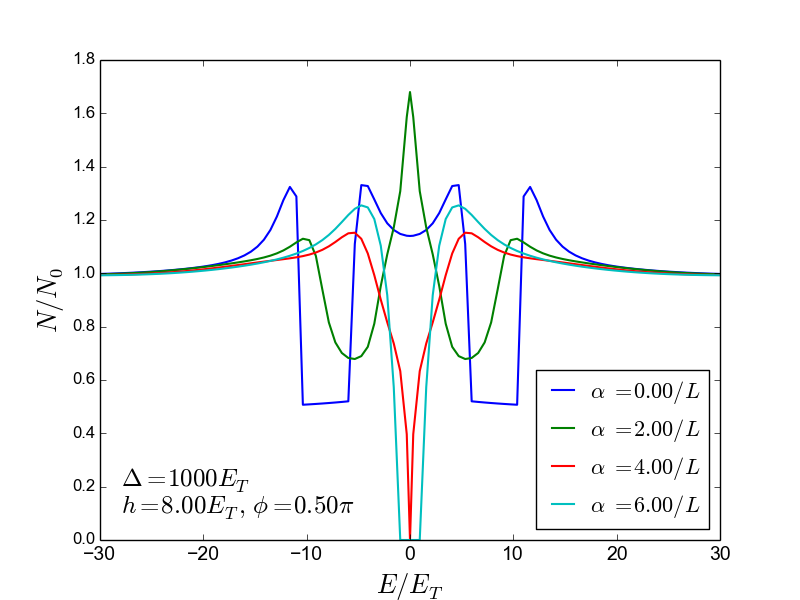}
\caption{Local density of states (DOS) in the center of the wire with
  $h = 8.00 E_T$ and $\phi = 0.50 \pi$ given with different spin-orbit
  fields. The DOS peaks at zero-energy for $\alpha = 2.00 / L$ while
  the peak converts to a zero-energy dip for larger  $\alpha$. \label{DOS-h-8.00}}
\end{figure}

In Fig.~\ref{DOS-h-8.00} we show the DOS peaking at zero-energy for
Rashba field or $\alpha = 2 / L$ while for larger $\alpha$ the peak
converts to a zero-energy dip and again to an energy gap centered
around zero energy. As discussed in Refs.~\onlinecite{tanaka07} and
\onlinecite{jacobsen15a}, the particular value for the zero-energy density of states results from
the competition of the singlet proximity effect aiming to lower $N(0)$
and the long-range triplet proximity effect to increase it. This
results from the different symmetry of the singlet vs. triplet
components. Namely, it is straightforward to show from the linearized
equations, Eq.~\eqref{eq:fjgeneral}, that for the singlet component $\tilde
f_2(\epsilon=0) = f_2^*(\epsilon=0)$, whereas the triplet components
satisfy $\tilde f_{j}(\epsilon=0)=-f_j^*(\epsilon=0)$, $j\neq 2$. Expanding
Eq.~\eqref{eq:ldos} to the lowest order in $\gamma$ then yields
\begin{equation}
N(0) \approx N_F \left[1-2|f_2(0)|^2+2\sum_{j\neq 2}
  |f_j(0)|^2\right].
\end{equation}
For large $h$, the singlet component vanishes on a short distance
$\sim \ell_m$ at the interface, whereas the long-range triplet
components $f_{0,3}$ are much larger in the center of the wire. As a
result, the latter yield $N(0)>1$. As shown in the figures, this may
signal the presence of a zero-energy peak, but not necessarily. The values
of $\alpha$ where the transition from a peak to a dip takes place are
similar to those yielding a maximum in the long-range supercurrent,
Fig.~\ref{fig:alpha_dependence}. 

\begin{figure}[h!]
\includegraphics[width=0.9\columnwidth]{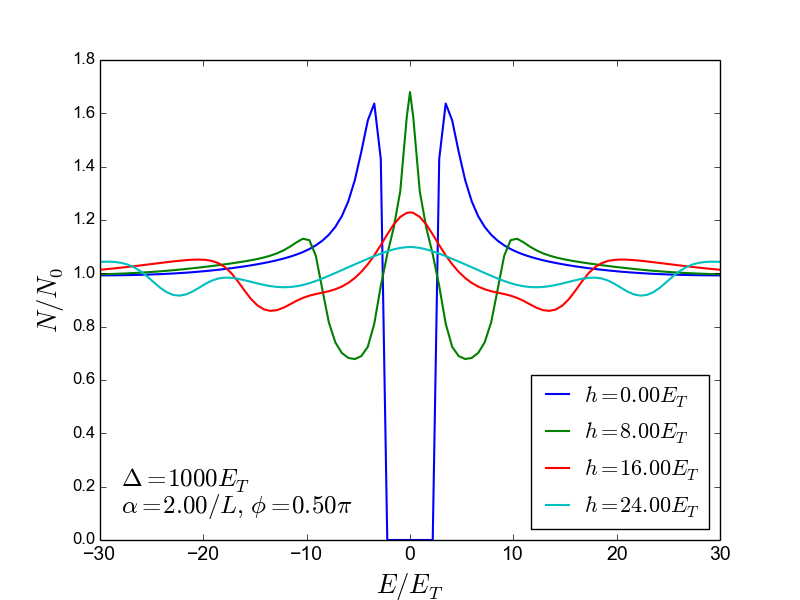}
\caption{Density of states with a constant $\alpha=2/L$ and with
  different values of the exchange field $h$. The zero-energy peak
  forms only for $h \gtrsim 2E_g(\phi)$. \label{DOS-alpha-2.00}}
\end{figure}

Although the origin is different from that expected for Majorana
junctions \cite{Science.336.1003,das12,churchill13}, also for the
diffusive nanowire Josephson junctions the zero-energy peak appears
only at a large enough exchange field, in practice for $h \gtrsim
2E_g(\phi)$ and for a restricted range of the values of $\alpha$. We plot the density of
states at a constant $\alpha$ for a few different exchange fields in
Fig.~\ref{DOS-alpha-2.00}. The exact form of the density of states
depends a lot on the value of the exchange fields, and the zero-bias
peak decays at large values of $h$.

\begin{figure}[h!]
\includegraphics[width=0.9\columnwidth]{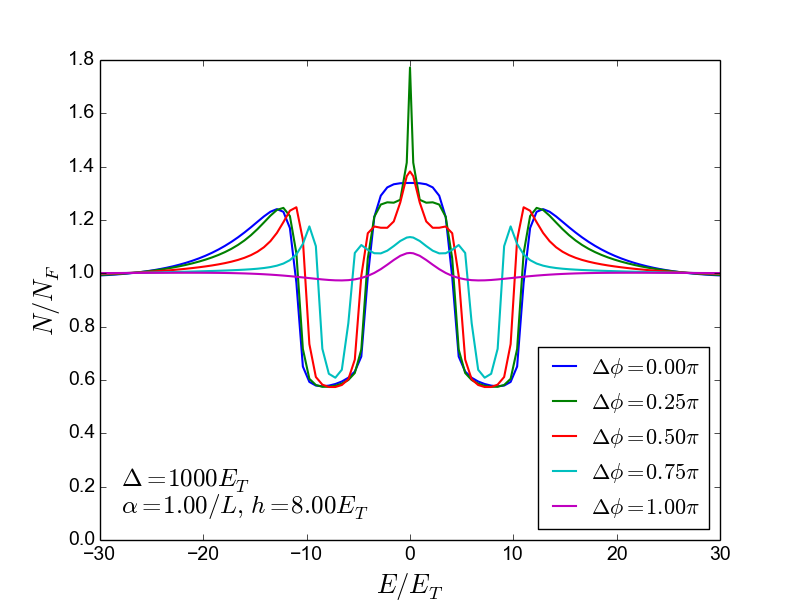}
\caption{Phase dependence of the DOS in the middle of the nanowire. \label{Phase-dependence-of-the-DOS-1.00-8.00}}
\end{figure}

\begin{figure}[h!]
\includegraphics[width=0.9\columnwidth]{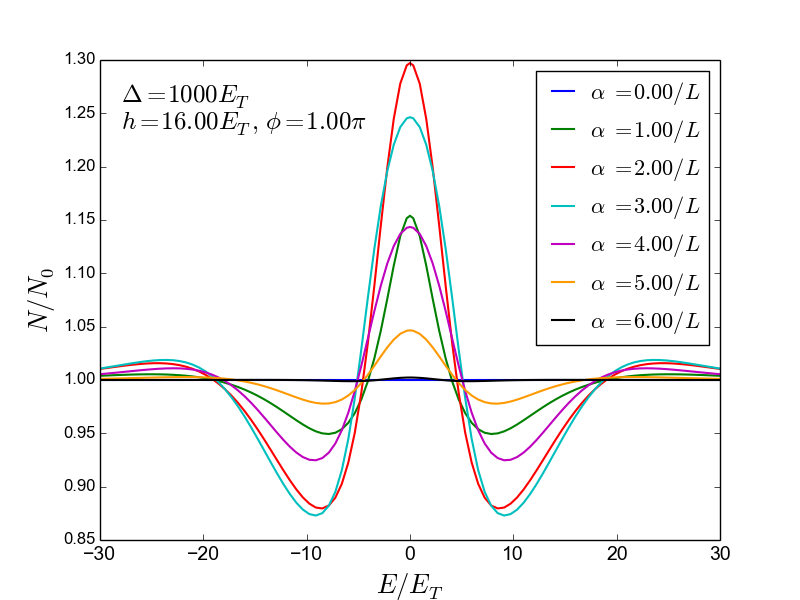}
\caption{Even with phase difference of $\phi = \pi$, the DOS peaks at zero energy for a finite SO coupling strength $\alpha$. \label{DOS-peak-phase-1.00-h-16.00}}
\end{figure}

The density of states can also be controlled by applying a
supercurrent through the junction, so that the phase difference $\phi$
across the junction changes. Besides changing (and closing) the
minigap for a vanishing exchange field, this changes the form of the
density of states. The phase dependence is plotted in
Fig.~\ref{Phase-dependence-of-the-DOS-1.00-8.00}. Note that typically
in long junctions ($\Delta \gg E_T$ or $L\gg \xi_0=\sqrt{\hbar D/\Delta}$)
the density of states for $\phi=\pi$ is almost featureless due to the
destructive interference of the pair amplitudes emanating from the two
superconductors. This is also modified by the spin-orbit coupling as
shown in Fig.~\ref{DOS-peak-phase-1.00-h-16.00}; also in this case a
zero-energy peak forms.

\begin{figure}[h]
\includegraphics[width=0.9\columnwidth]{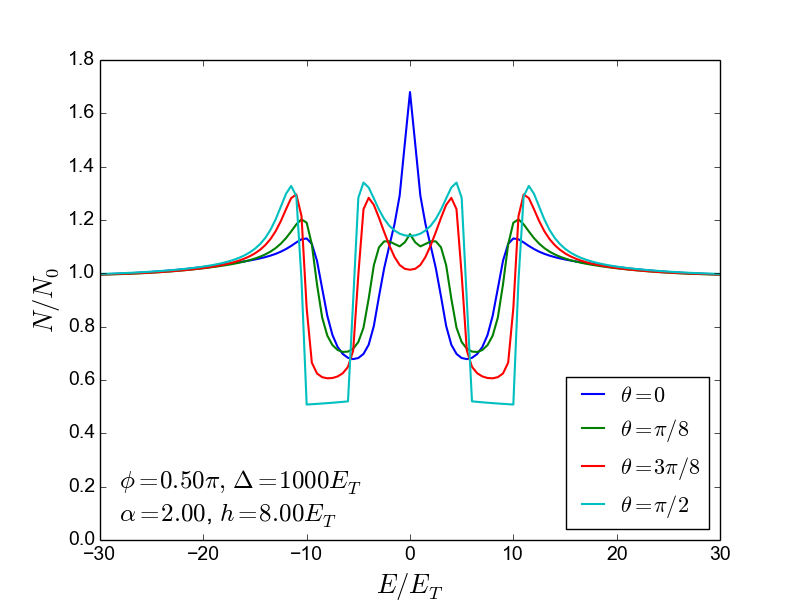}
\caption{Density of states for different directions of the exchange
  field, specified in terms of the angle $\theta$ (see inset of Fig.~\ref{fig:supercurrent_exchange_field_angular_dependence}).}
\label{fig:DOS-vs-angle}
\end{figure}

Let us furthermore demonstrate the connection between the triplet
proximity effect and the zero-energy peak.\cite{konstandin05,alidoust10,jacobsen15b,alidoust15} Therefore, we study the density of states as a function of the
angle $\theta$ of the exchange field between the $z$ direction of the
wire and the $x$ direction, as in Fig.~\ref{fig:supercurrent_exchange_field_angular_dependence} for the supercurrent. This is
shown in Fig.~\ref{fig:DOS-vs-angle}. For
$\theta=\pi/2$ we expect to get only the short-range proximity effect. In this case
we indeed only find the exchange-field split minigaps, and the
spin-orbit coupling makes almost no contribution to the form of the
density of states. 

\section{Conclusions}
In this work we have discussed the detailed effects of the
Rashba/Dresselhaus -type intrinsic spin-orbit interaction on the
supercurrent carried through a diffusive nanowire. We solve the full
Usadel equation to obtain the supercurrent of the junction and the
local density of states. We reproduce the long-range triplet proximity
effect predicted before in the linearized limit and show how the
resulting supercurrent depends on the direction of the applied
field. Besides the complicated direction dependence and the predicted
zero-energy peaks and dips in the local density of states, our results
pave the way of quantitatively analyzing the experiments carried out
in the nanowire Josephson junctions. To reach for example the
topological regime and confirm the Majorana character of the
excitations in the nanowires, it is important that the
experimentalists are able to characterize their junctions in
detail. Our work gives a fixed point for such a characterization, in
the (non-topological) many-channel diffusive limit. We hence expect
this work to be relevant as an intermediate step for establishing the
experimental constraints for using such junctions in topological
quantum computing.

We thank Pauli Virtanen, Sebastian Bergeret and Timo Hyart for very useful
discussions, and Francesco Giazotto, Attila Geresdi and Charles Marcus
for explaining the characteristics of InAs/InSb nanowires. This work was supported by the Academy of Finland through its Center of Excellence program, and by the European Research Council (Grant No. 240362-Heattronics).

\appendix
\section{Riccati parameterization} \label{sec:RiccatiAppendix}

The normalization condition $\left(\hat{G}^{R}\right)^{2}=1$ implies that the possible eigenvalues of $\hat{G}^{R}$ are $\pm1$. Therefore in spectral representation, $\hat{G}^{R}$ can be written in terms of so called Shelankov projectors as \cite{J.Low.Temp.Phys.60.29}

\begin{equation}
\hat{G}^{R}=\hat{P}_{+} - \hat{P}_{-} \quad\text{with} \quad \hat{P}_{\pm} = \frac{1}{2}\left(1\pm\hat{G}^{R}\right). \label{ShelankovProjectors}
\end{equation}

The Shelankov projectors and the Green's function are convenient to parametrize in the Riccati parameterization \cite{Phys.Rev.B.61.9061}

\begin{equation}
\hat{P}_{+} = 
\begin{pmatrix}
N			& N \gamma 			\\
\tilde{\gamma} N	& \tilde{\gamma} N \gamma
\end{pmatrix}
\quad \text{and} \quad
\hat{P}_{-} = 
\begin{pmatrix}
\gamma \tilde{N} \tilde{\gamma} & -\gamma \tilde{N} \\
-\tilde{N} \tilde{\gamma}	& \tilde{N}
\end{pmatrix}, \\
\end{equation}

\noindent where $N = \left( 1 + \gamma \tilde{\gamma} \right)^{-1}$
and $\tilde{N} = \left( 1 + \tilde{\gamma} \gamma \right)^{-1}$. Thus
from Eq. (\ref{ShelankovProjectors}) the Green's function is

\begin{equation}
\hat{G}^R = 
\begin{pmatrix}
N	&	0			\\
0	&	\tilde{N}
\end{pmatrix}
\begin{pmatrix}
1 - \gamma \tilde{\gamma}	&	2 \gamma			\\
2 \tilde{\gamma}		&	-(1 - \tilde{\gamma} \gamma)
\end{pmatrix}.
\end{equation}

The projectors have the property

\begin{equation}
\hat{\nabla}\hat{P}_{\pm} = \pm\hat{P}_{+}\left[\hat{\nabla}U\right]\hat{P}_{-}\pm\hat{P}_{-}\left[\hat{\nabla}\tilde{U}\right]\hat{P}_{+},
\end{equation}

where

\begin{equation}
U = 
\begin{pmatrix}
0	&	\gamma	\\
0	&	0
\end{pmatrix}
\quad \text{and} \quad
\tilde{U} = 
\begin{pmatrix}
0		&	0	\\
\tilde{\gamma}	&	0	\\
\end{pmatrix}.
\end{equation}

In the spin-dependent case, $\gamma$ and $\tilde{\gamma}$ are $2 \times 2$-spin matrices, that is 

\begin{equation}
\gamma = \sum_{j=0}^{3} \gamma_j \bar\sigma_j \quad \text{and} \quad \tilde{\gamma} = \sum_{j=0}^{3} \tilde{\gamma}_j \bar\sigma_j.
\end{equation}

Using the above relations, we have derived the Usadel equation \eqref{UsadelRiccati}
and the spectral supercurrent, Eq.~\eqref{spectral_current_density_riccati}.

\section{Linearized equations}

Linearizing Eq.~\eqref{UsadelRiccati} yields 

\begin{equation}
\gamma'' = 2 \omega_n \gamma + i [\gamma, \bar{h}] + [\bar{\textbf{A}}_s^2, \gamma] + 2 \{\bar{\textbf{A}}_s, \gamma\}\bar{\textbf{A}}_s + 2i \{\gamma', \bar{\textbf{A}}_s\},
\end{equation}

\noindent where we use a shorthand notation $\partial_{z'} \gamma = \gamma'$. Writing 

\begin{equation}
\gamma = \sum_{i=0}^{3} f_i \bar\sigma_i \label{pauli_matrix_expansion_of_gamma},
\end{equation}

\noindent assuming a general form for the SO coupling 

\begin{equation}
\bar{\mathbf{A}}=\begin{pmatrix} \alpha_1 \bar{\sigma}_1 + \alpha_2 \bar{\sigma}_3, & 0, &
  \alpha_3 \bar{\sigma}_1 + \alpha_4 \bar{\sigma}_3 \end{pmatrix}. \label{general_rashba}
\end{equation}

\noindent and choosing the exchange field in plane

\begin{equation}
\bar{h} = h (\text{cos}\theta \bar{\sigma}_3 + \text{sin}\theta \bar{\sigma}_1) \label{exchange_with_angle}
\end{equation}

\noindent we can separate the equations for the different spin
components. They read

\begin{subequations}
\label{eq:fjgeneral}
\begin{align*}
f_0'' &= (2 \omega_n + 4 \sum_{i=1}^{4} \alpha_i^2) f_0 + 4 i \alpha_3 f_1' + 4 i \alpha_4 f_3' \addtag \label{f_0_general_appendix} \\
f_1'' &= (2 \omega_n + 4(\alpha_1^2 + \alpha_ 3^2)) f_1 - 2h \text{cos}\theta f_2 \\
&+ 4(\alpha_1 \alpha_2 + \alpha_3 \alpha_4) f_3 + 4 i \alpha_3 f_0' \addtag \label{f_1_general_appendix} \\
f_2'' &= 2h \text{cos}\theta f_1 + 2 \omega_n f_2 - 2h \text{sin}\theta f_3 \addtag \label{f_2_general_appendix} \\ 
f_3'' &= 4(\alpha_1 \alpha_2 + \alpha_3 \alpha_4) f_1 + 2h \text{sin}\theta f_2\\
&+ (2 \omega_n + 4 (\alpha_2^2 + \alpha_4^2)) f_3. \addtag \label{f_3_general_appendix}
\end{align*}
\end{subequations}

\noindent and the equations for $\tilde{f}_i$ parametrizing $\tilde
\gamma$ are obtained by substituting $f_i \leftrightarrow \tilde{f}_i$, $h \leftrightarrow -h$, and by taking a complex conjugate of the scalars.

The boundary conditions read 

\begin{align}
f_2 = c e^{i\phi/2}, \tilde{f}_2 = c e^{-i\phi/2}, c= \frac{|\Delta|}{\omega_n + \sqrt{|\Delta|^2 + \omega_n^2}}, \label{f_2_bound_appendix}
\end{align}

\noindent with $\phi$ having the opposite signs at $x = 0$ and $x = L$. For $i=0,1,3$ the functions vanish at the boundaries, that is $f_i(0) = f_i(L) = \tilde{f}_i(0) = \tilde{f}_i(L) = 0$.

Linearizing the spectral current density (see
Eq.~\eqref{spectral_current_density_riccati}) yields

\begin{align*}
&{\rm Im} (j_{s}) = \frac{1}{2}\text{Im}\left\{ \text{Tr}\left[ \left(\gamma\tilde{\gamma}'-\gamma'\tilde{\gamma}\right)-\left(\tilde{\gamma}\gamma'-\tilde{\gamma}'\gamma\right)\right.\right. \addtag \label{spectral_current_density_linearized}\\
&\left.\left.+i\bigg(\left\{ \bar{\textbf{A}},\gamma\right\} \tilde{\gamma}+\gamma\left\{ \bar{\textbf{A}},\tilde{\gamma}\right\} \right.\right.
\left.\left.+\left\{ \bar{\textbf{A}},\tilde{\gamma}\right\} \gamma + \tilde{\gamma}\left\{ \bar{\textbf{A}},\gamma\right\} \bigg)\right]\right\}. 
\end{align*}

\noindent Using the Pauli matrix expansion from Eq.~\eqref{pauli_matrix_expansion_of_gamma} and the general form for the SO coupling, Eq.~\eqref{general_rashba} we can simplify Eq.~\eqref{spectral_current_density_linearized} 

\begin{align*}
{\rm Im} (j_{s}) &= \text{Im} \bigg[ \sum_{j=0}^{3} (f_j \tilde{f}'_j - \tilde{f}_j f'_j) \\
&+ 4 i ( \alpha_3 (f_0 \tilde{f}_1 + f_1 \tilde{f}_0) + \alpha_4 (f_0 \tilde{f}_3 + f_3 \tilde{f}_0) )\bigg]. \addtag
\end{align*}
In the numerics we have chosen $\bar{\textbf{A}} = (0, 0, -\alpha
\bar{\sigma}_1)$. On the other hand, the analytics becomes more
straightforward by writing the $f$-function components in the basis
dictated by the exchange field. Therefore, applying the rotation
$\exp(i\theta \sigma_y/2)$ to $\mathbf{\tilde h}$, $\gamma$ and
$\bar{\textbf{A}}$ yields 

\begin{align*}
f_0'' =& (2 \omega_n + 4 \alpha^2) f_0 - 4 i \alpha \cos(\theta) f_1'
        -4i\alpha \sin(\theta) f_3'\\
f_1'' =& [2 \omega_n + 4 \alpha^2 \cos^2(\theta)] f_1 - 2 h f_2 - 4 i
        \alpha \cos(\theta) f_0' \\&+2\alpha^2
        \sin(2\theta)f_3\\
f_2'' =& 2 h f_1 + 2 \omega_n f_2 \\ 
f_3'' =& [2 \omega_n + 4 \alpha^2 \sin^2(\theta)] f_3+2 \alpha^2
        \sin(2\theta)f_1-4 i \alpha \sin(\theta) f_0' .
\end{align*}

The corresponding equations for $\tilde f_j$ are obtained after
replacing $h \leftrightarrow -h$ and $\alpha \leftrightarrow
-\alpha$. 

For $h \gg \omega_n,\alpha$, it is now straightforward to identify $f_1$ and $f_2$ as the
short-range components, decaying within the length $\xi_m =
1/\sqrt{2h}$ from the interfaces. These two components are the $m=0$
triplet and the singlet component of the pairing amplitude,
respectively. These components can thus be solved separately in a
straightforward manner, but the resulting analytic expressions are too
lengthy to be printed here. 

The short-range components generate boundary conditions for the
long-range components $f_0$ and $f_3$ that decay within the length
$\ell_0 = 1/\sqrt{2\omega_n + 4 \alpha^2)}, \ell_3 = 1/\sqrt{2 \omega_n
    + 4 \alpha^2 \sin(\theta)}\gg \xi_m$. First
  disregarding the coupling term between $f_0$ and $f_3$, we get an
  analytic solution,
\begin{align}
f_0 &= A_0 \sinh\left[\frac{z}{\ell_0}\right]+4 i \alpha \cos(\theta)  \int_0^z \ell_0 
  \sinh\left[\frac{z-x}{\ell_0}\right] f_1'(x) dx\\
f_3 &= A_3 \sinh\left[\frac{z}{\ell_3}\right]+2 \alpha^2 \sin(2\theta)  \int_0^z \ell_3 
  \sinh\left[\frac{z-x}{\ell_3}\right] f_1(x) dx.
\end{align}
This solution takes into account the boundary condition
$f_0(0)=f_3(0)=0$. The prefactors $A_{0,3}$ would be obtained from the
other boundary condition $f_0(1)=f_3(0)=1$. However, we concentrate
only on the vicinity of the interface at $z=0$, and disregard these
terms. Note that both components vanish for $\theta=\pi/2$, when the
exchange field is collinear with the spin-orbit field, whereas $f_3=0$
for $\theta=0$. 

For $\ell_0 \gg z \gg \xi_m$, we may expand and get
\begin{align}
f_0(z) &\approx 4 i \alpha \cos(\theta) \int_0^z f_1(x)dx\\
f_3'(z) &\approx 2 \alpha^2 \sin(2\theta) \int_0^z f_1(x)dx.
\end{align}
The previous expression requires also one partial integration and
using the fact that $f_1(0)=0$. The equation is written for $f_3'(z)$
because only the derivative tends to a constant in this
interval. Performing the integral yields $\int_0^z f_1(x) \approx c
\exp(i\phi/2)/(2\sqrt{h})$, saturating for $z \gg \xi_m$. Repeating a similar procedure on the other
end, $z=1$ (in reduced units) yields finally the full boundary
conditions for the long-range components
\begin{align}
f_0(0)&=2 i \alpha \cos(\theta) c e^{i\phi/2}/\sqrt{h}\\
f_3'(0)&=\alpha^2 \sin(2\theta) c e^{i\phi/2}/\sqrt{h}\\
f_0(1)&=2 i \alpha \cos(\theta) c e^{-i\phi/2}/\sqrt{h}\\
f_3'(1)&=-\alpha^2 \sin(2\theta) c e^{-i\phi/2}/\sqrt{h},
\end{align}
where now $z=0,1$ mean the position $\sim \xi_m$ away from the
contacts. The boundary conditions for $\tilde f_{0,3}$ are obtained by
changing the sign of the derivatives, $\alpha \leftrightarrow
-\alpha$, $\phi \leftrightarrow -\phi$. The remaining equations for the long-range components can be
written for the two-component vector $\vec f = \begin{pmatrix} f_0 &
  f_3 \end{pmatrix}^T$ as 
\begin{equation}
\vec f'' = [\omega_n + 4 \alpha^2] \vec{f} - 4 \alpha^2
\cos^2(\theta) \sigma_\downarrow \vec{f} - 4 i \alpha \sin(\theta)
\sigma_x \vec{f}'.
\end{equation}
We get rid of the second off-diagonal term by defining $\vec{h} =
U(z)\vec{f} = \exp(-2 i \alpha \sin(\theta) \sigma_x z) \vec{f}
$. This satisfies
\begin{equation}
\vec h''=(\omega_n + 4 \alpha^2 \cos^2(\theta))\vec{h} - 4 \alpha^2 \cos^2(\theta) U(z)
\sigma_\downarrow U^\dagger(z) \vec{h}\label{eq:heq}
\end{equation}
with $\vec{h}^{(\prime)}(0)=\vec{f}(0)$,
$\exp(2i\alpha\sin(\theta)\sigma_x)
\vec{h}^{(\prime)}(1)=\vec{f}(1)$. Besides Eq.~\eqref{eq:heq} we can
hence transform the coupling terms to the boundary conditions. As the
general solution to Eq.~\eqref{eq:heq} is lengthy, we first disregard
this term. In this case, solving \eqref{eq:heq} is
straightforward. The full solution for $\vec{h}(z)$ is lengthy,
but the spectral supercurrent is given by $j_s=8i\alpha^2
\cos^2(\theta) c^2
\sin(\phi) I(\alpha)$ with
\begin{widetext}
\begin{equation}
I(\alpha)= \frac{ e^{\sqrt{2} \sqrt{\tilde \omega_n}} \left(\sqrt{2} \left(e^{2 \sqrt{2} \sqrt{\tilde \omega_n}}-1\right) \cos (2 \alpha  \sin (\theta )) \left(2 \tilde \omega_n-\alpha ^2 \sin
   ^2(\theta )\right)-4 \alpha  \sin (\theta ) \left(e^{2 \sqrt{2} \sqrt{\tilde \omega_n}}+1\right) \sqrt{\tilde \omega_n} \sin (2 \alpha  \sin (\theta ))\right)}{h \sqrt{\tilde \omega_n} \left(-2 e^{2 \sqrt{2} \sqrt{\tilde \omega_n}} \cos (4
   \alpha  \sin (\theta ))+e^{4 \sqrt{2} \sqrt{\tilde
       \omega_n}}+1\right)},
\end{equation}
\end{widetext}
where $\tilde \omega_n = \omega_n + 2 \alpha^2 \cos^2(\theta)$. In
particular, for $\theta=0$, where the above approximation of
neglecting the $\sigma_\downarrow$ term is not
relevant, we get Eq.~\eqref{eq:spectralscanalytic}.

For $\theta$ between 0 and $\pi/2$, the two long-range components mix
and produce a supercurrent that can change sign for a fixed
phase as the direction of the magnetic field is tuned. Such a sign change is described by the terms of the form
$\cos(2 \alpha \sin(\theta))$ that result from the above
transformation between $\vec{f}$ and $\vec{h}$. They show that the
sign change takes place whenever $\alpha \gtrsim 1$. However, due to
the approximation of neglecting the $\sigma_\downarrow$ term, this
result does not match very well the exact solution at arbitrary range
of parameters. Therefore, it is better to use the rather
straightforward solution of the full linearized equations for finding
the supercurrent in this case.

\bibliography{Superconductivity.bib}

\begin{thebibliography}{54}%
\makeatletter
\providecommand \@ifxundefined [1]{%
 \@ifx{#1\undefined}
}%
\providecommand \@ifnum [1]{%
 \ifnum #1\expandafter \@firstoftwo
 \else \expandafter \@secondoftwo
 \fi
}%
\providecommand \@ifx [1]{%
 \ifx #1\expandafter \@firstoftwo
 \else \expandafter \@secondoftwo
 \fi
}%
\providecommand \natexlab [1]{#1}%
\providecommand \enquote  [1]{``#1''}%
\providecommand \bibnamefont  [1]{#1}%
\providecommand \bibfnamefont [1]{#1}%
\providecommand \citenamefont [1]{#1}%
\providecommand \href@noop [0]{\@secondoftwo}%
\providecommand \href [0]{\begingroup \@sanitize@url \@href}%
\providecommand \@href[1]{\@@startlink{#1}\@@href}%
\providecommand \@@href[1]{\endgroup#1\@@endlink}%
\providecommand \@sanitize@url [0]{\catcode `\\12\catcode `\$12\catcode
  `\&12\catcode `\#12\catcode `\^12\catcode `\_12\catcode `\%12\relax}%
\providecommand \@@startlink[1]{}%
\providecommand \@@endlink[0]{}%
\providecommand \url  [0]{\begingroup\@sanitize@url \@url }%
\providecommand \@url [1]{\endgroup\@href {#1}{\urlprefix }}%
\providecommand \urlprefix  [0]{URL }%
\providecommand \Eprint [0]{\href }%
\providecommand \doibase [0]{http://dx.doi.org/}%
\providecommand \selectlanguage [0]{\@gobble}%
\providecommand \bibinfo  [0]{\@secondoftwo}%
\providecommand \bibfield  [0]{\@secondoftwo}%
\providecommand \translation [1]{[#1]}%
\providecommand \BibitemOpen [0]{}%
\providecommand \bibitemStop [0]{}%
\providecommand \bibitemNoStop [0]{.\EOS\space}%
\providecommand \EOS [0]{\spacefactor3000\relax}%
\providecommand \BibitemShut  [1]{\csname bibitem#1\endcsname}%
\let\auto@bib@innerbib\@empty
%</preamble>
\bibitem [{num()}]{numericsnote}%
  \BibitemOpen
  \href@noop {} {}\bibinfo {note} {The code used for calculating the results in
  this paper will be made available upon the publication of this manuscript and
  before that can be requested from the authors.}\BibitemShut {Stop}%
\bibitem [{\citenamefont {Ryazanov}\ \emph {et~al.}(2001)\citenamefont
  {Ryazanov}, \citenamefont {Oboznov}, \citenamefont {Rusanov}, \citenamefont
  {Veretennikov}, \citenamefont {Golubov},\ and\ \citenamefont
  {Aarts}}]{ryazanov01}%
  \BibitemOpen
  \bibfield  {author} {\bibinfo {author} {\bibfnamefont {V.~V.}\ \bibnamefont
  {Ryazanov}}, \bibinfo {author} {\bibfnamefont {V.~A.}\ \bibnamefont
  {Oboznov}}, \bibinfo {author} {\bibfnamefont {A.~Y.}\ \bibnamefont
  {Rusanov}}, \bibinfo {author} {\bibfnamefont {A.~V.}\ \bibnamefont
  {Veretennikov}}, \bibinfo {author} {\bibfnamefont {A.~A.}\ \bibnamefont
  {Golubov}}, \ and\ \bibinfo {author} {\bibfnamefont {J.}~\bibnamefont
  {Aarts}},\ }\href {\doibase 10.1103/PhysRevLett.86.2427} {\bibfield
  {journal} {\bibinfo  {journal} {Phys. Rev. Lett.}\ }\textbf {\bibinfo
  {volume} {86}},\ \bibinfo {pages} {2427} (\bibinfo {year}
  {2001})}\BibitemShut {NoStop}%
\bibitem [{\citenamefont {Kontos}\ \emph {et~al.}(2001)\citenamefont {Kontos},
  \citenamefont {Aprili}, \citenamefont {Lesueur},\ and\ \citenamefont
  {Grison}}]{kontos01}%
  \BibitemOpen
  \bibfield  {author} {\bibinfo {author} {\bibfnamefont {T.}~\bibnamefont
  {Kontos}}, \bibinfo {author} {\bibfnamefont {M.}~\bibnamefont {Aprili}},
  \bibinfo {author} {\bibfnamefont {J.}~\bibnamefont {Lesueur}}, \ and\
  \bibinfo {author} {\bibfnamefont {X.}~\bibnamefont {Grison}},\ }\href
  {\doibase 10.1103/PhysRevLett.86.304} {\bibfield  {journal} {\bibinfo
  {journal} {Phys. Rev. Lett.}\ }\textbf {\bibinfo {volume} {86}},\ \bibinfo
  {pages} {304} (\bibinfo {year} {2001})}\BibitemShut {NoStop}%
\bibitem [{\citenamefont {Bergeret}\ \emph {et~al.}(2001)\citenamefont
  {Bergeret}, \citenamefont {Volkov},\ and\ \citenamefont
  {Efetov}}]{bergeret01}%
  \BibitemOpen
  \bibfield  {author} {\bibinfo {author} {\bibfnamefont {F.~S.}\ \bibnamefont
  {Bergeret}}, \bibinfo {author} {\bibfnamefont {A.~F.}\ \bibnamefont
  {Volkov}}, \ and\ \bibinfo {author} {\bibfnamefont {K.~B.}\ \bibnamefont
  {Efetov}},\ }\href {\doibase 10.1103/PhysRevLett.86.4096} {\bibfield
  {journal} {\bibinfo  {journal} {Phys. Rev. Lett.}\ }\textbf {\bibinfo
  {volume} {86}},\ \bibinfo {pages} {4096} (\bibinfo {year}
  {2001})}\BibitemShut {NoStop}%
\bibitem [{\citenamefont {Khaire}\ \emph {et~al.}(2010)\citenamefont {Khaire},
  \citenamefont {Khasawneh}, \citenamefont {Pratt},\ and\ \citenamefont
  {Birge}}]{khaire10}%
  \BibitemOpen
  \bibfield  {author} {\bibinfo {author} {\bibfnamefont {T.~S.}\ \bibnamefont
  {Khaire}}, \bibinfo {author} {\bibfnamefont {M.~A.}\ \bibnamefont
  {Khasawneh}}, \bibinfo {author} {\bibfnamefont {W.~P.}\ \bibnamefont
  {Pratt}}, \ and\ \bibinfo {author} {\bibfnamefont {N.~O.}\ \bibnamefont
  {Birge}},\ }\href {\doibase 10.1103/PhysRevLett.104.137002} {\bibfield
  {journal} {\bibinfo  {journal} {Phys. Rev. Lett.}\ }\textbf {\bibinfo
  {volume} {104}},\ \bibinfo {pages} {137002} (\bibinfo {year}
  {2010})}\BibitemShut {NoStop}%
\bibitem [{\citenamefont {Robinson}\ \emph {et~al.}(2010)\citenamefont
  {Robinson}, \citenamefont {Witt},\ and\ \citenamefont
  {Blamire}}]{robinson10}%
  \BibitemOpen
  \bibfield  {author} {\bibinfo {author} {\bibfnamefont {J.~W.~A.}\
  \bibnamefont {Robinson}}, \bibinfo {author} {\bibfnamefont {J.~D.~S.}\
  \bibnamefont {Witt}}, \ and\ \bibinfo {author} {\bibfnamefont {M.~G.}\
  \bibnamefont {Blamire}},\ }\href {\doibase 10.1126/science.1189246}
  {\bibfield  {journal} {\bibinfo  {journal} {Science}\ }\textbf {\bibinfo
  {volume} {329}},\ \bibinfo {pages} {59} (\bibinfo {year} {2010})}\BibitemShut
  {NoStop}%
\bibitem [{\citenamefont {Heikkil\"a}\ \emph {et~al.}(2000)\citenamefont
  {Heikkil\"a}, \citenamefont {Wilhelm},\ and\ \citenamefont
  {Sch\"on}}]{heikkila00}%
  \BibitemOpen
  \bibfield  {author} {\bibinfo {author} {\bibfnamefont {T.~T.}\ \bibnamefont
  {Heikkil\"a}}, \bibinfo {author} {\bibfnamefont {F.~K.}\ \bibnamefont
  {Wilhelm}}, \ and\ \bibinfo {author} {\bibfnamefont {G.}~\bibnamefont
  {Sch\"on}},\ }\href {\doibase 10.1209/epl/i2000-00513-x} {\bibfield
  {journal} {\bibinfo  {journal} {Europhys. Lett.}\ }\textbf {\bibinfo {volume}
  {51}},\ \bibinfo {pages} {434} (\bibinfo {year} {2000})}\BibitemShut
  {NoStop}%
\bibitem [{\citenamefont {Yip}(2000)}]{yip00}%
  \BibitemOpen
  \bibfield  {author} {\bibinfo {author} {\bibfnamefont {S.-K.}\ \bibnamefont
  {Yip}},\ }\href {\doibase 10.1103/PhysRevB.62.R6127} {\bibfield  {journal}
  {\bibinfo  {journal} {Phys. Rev. B}\ }\textbf {\bibinfo {volume} {62}},\
  \bibinfo {pages} {R6127} (\bibinfo {year} {2000})}\BibitemShut {NoStop}%
\bibitem [{\citenamefont {Crosser}\ \emph {et~al.}(2008)\citenamefont
  {Crosser}, \citenamefont {Huang}, \citenamefont {Pierre}, \citenamefont
  {Virtanen}, \citenamefont {Heikkil\"a}, \citenamefont {Wilhelm},\ and\
  \citenamefont {Birge}}]{crosser08}%
  \BibitemOpen
  \bibfield  {author} {\bibinfo {author} {\bibfnamefont {M.~S.}\ \bibnamefont
  {Crosser}}, \bibinfo {author} {\bibfnamefont {J.}~\bibnamefont {Huang}},
  \bibinfo {author} {\bibfnamefont {F.}~\bibnamefont {Pierre}}, \bibinfo
  {author} {\bibfnamefont {P.}~\bibnamefont {Virtanen}}, \bibinfo {author}
  {\bibfnamefont {T.~T.}\ \bibnamefont {Heikkil\"a}}, \bibinfo {author}
  {\bibfnamefont {F.~K.}\ \bibnamefont {Wilhelm}}, \ and\ \bibinfo {author}
  {\bibfnamefont {N.~O.}\ \bibnamefont {Birge}},\ }\href {\doibase
  10.1103/PhysRevB.77.014528} {\bibfield  {journal} {\bibinfo  {journal} {Phys.
  Rev. B}\ }\textbf {\bibinfo {volume} {77}},\ \bibinfo {pages} {014528}
  (\bibinfo {year} {2008})}\BibitemShut {NoStop}%
\bibitem [{\citenamefont {Cuevas}\ and\ \citenamefont
  {Bergeret}(2007)}]{cuevas07}%
  \BibitemOpen
  \bibfield  {author} {\bibinfo {author} {\bibfnamefont {J.~C.}\ \bibnamefont
  {Cuevas}}\ and\ \bibinfo {author} {\bibfnamefont {F.~S.}\ \bibnamefont
  {Bergeret}},\ }\href {\doibase 10.1103/PhysRevLett.99.217002} {\bibfield
  {journal} {\bibinfo  {journal} {Phys. Rev. Lett.}\ }\textbf {\bibinfo
  {volume} {99}},\ \bibinfo {pages} {217002} (\bibinfo {year}
  {2007})}\BibitemShut {NoStop}%
\bibitem [{\citenamefont {Mourik}\ \emph {et~al.}(2012)\citenamefont {Mourik},
  \citenamefont {Zuo}, \citenamefont {Frolov}, \citenamefont {Plissard},
  \citenamefont {Bakkers},\ and\ \citenamefont
  {Kouwenhoven}}]{Science.336.1003}%
  \BibitemOpen
  \bibfield  {author} {\bibinfo {author} {\bibfnamefont {V.}~\bibnamefont
  {Mourik}}, \bibinfo {author} {\bibfnamefont {K.}~\bibnamefont {Zuo}},
  \bibinfo {author} {\bibfnamefont {S.~M.}\ \bibnamefont {Frolov}}, \bibinfo
  {author} {\bibfnamefont {S.~R.}\ \bibnamefont {Plissard}}, \bibinfo {author}
  {\bibfnamefont {E.~P. A.~M.}\ \bibnamefont {Bakkers}}, \ and\ \bibinfo
  {author} {\bibfnamefont {L.~P.}\ \bibnamefont {Kouwenhoven}},\ }\href@noop {}
  {\bibfield  {journal} {\bibinfo  {journal} {Science}\ }\textbf {\bibinfo
  {volume} {336}},\ \bibinfo {pages} {1003} (\bibinfo {year}
  {2012})}\BibitemShut {NoStop}%
\bibitem [{\citenamefont {Das}\ \emph {et~al.}(2012)\citenamefont {Das},
  \citenamefont {Ronen}, \citenamefont {Most}, \citenamefont {Oreg},
  \citenamefont {Heiblum},\ and\ \citenamefont {Shtrikman}}]{das12}%
  \BibitemOpen
  \bibfield  {author} {\bibinfo {author} {\bibfnamefont {A.}~\bibnamefont
  {Das}}, \bibinfo {author} {\bibfnamefont {Y.}~\bibnamefont {Ronen}}, \bibinfo
  {author} {\bibfnamefont {Y.}~\bibnamefont {Most}}, \bibinfo {author}
  {\bibfnamefont {Y.}~\bibnamefont {Oreg}}, \bibinfo {author} {\bibfnamefont
  {M.}~\bibnamefont {Heiblum}}, \ and\ \bibinfo {author} {\bibfnamefont
  {H.}~\bibnamefont {Shtrikman}},\ }\href@noop {} {\bibfield  {journal}
  {\bibinfo  {journal} {Nature Phys.}\ }\textbf {\bibinfo {volume} {8}},\
  \bibinfo {pages} {887} (\bibinfo {year} {2012})}\BibitemShut {NoStop}%
\bibitem [{\citenamefont {Churchill}\ \emph {et~al.}(2013)\citenamefont
  {Churchill}, \citenamefont {Fatemi}, \citenamefont {Grove-Rasmussen},
  \citenamefont {Deng}, \citenamefont {Caroff}, \citenamefont {Xu},\ and\
  \citenamefont {Marcus}}]{churchill13}%
  \BibitemOpen
  \bibfield  {author} {\bibinfo {author} {\bibfnamefont {H.~O.~H.}\
  \bibnamefont {Churchill}}, \bibinfo {author} {\bibfnamefont {V.}~\bibnamefont
  {Fatemi}}, \bibinfo {author} {\bibfnamefont {K.}~\bibnamefont
  {Grove-Rasmussen}}, \bibinfo {author} {\bibfnamefont {M.~T.}\ \bibnamefont
  {Deng}}, \bibinfo {author} {\bibfnamefont {P.}~\bibnamefont {Caroff}},
  \bibinfo {author} {\bibfnamefont {H.~Q.}\ \bibnamefont {Xu}}, \ and\ \bibinfo
  {author} {\bibfnamefont {C.~M.}\ \bibnamefont {Marcus}},\ }\href {\doibase
  10.1103/PhysRevB.87.241401} {\bibfield  {journal} {\bibinfo  {journal} {Phys.
  Rev. B}\ }\textbf {\bibinfo {volume} {87}},\ \bibinfo {pages} {241401}
  (\bibinfo {year} {2013})}\BibitemShut {NoStop}%
\bibitem [{\citenamefont {Sau}\ \emph {et~al.}(2010)\citenamefont {Sau},
  \citenamefont {Lutchyn}, \citenamefont {Tewari},\ and\ \citenamefont
  {Das~Sarma}}]{PhysRevLett.104.040502}%
  \BibitemOpen
  \bibfield  {author} {\bibinfo {author} {\bibfnamefont {J.~D.}\ \bibnamefont
  {Sau}}, \bibinfo {author} {\bibfnamefont {R.~M.}\ \bibnamefont {Lutchyn}},
  \bibinfo {author} {\bibfnamefont {S.}~\bibnamefont {Tewari}}, \ and\ \bibinfo
  {author} {\bibfnamefont {S.}~\bibnamefont {Das~Sarma}},\ }\href@noop {}
  {\bibfield  {journal} {\bibinfo  {journal} {Phys. Rev. Lett.}\ }\textbf
  {\bibinfo {volume} {104}},\ \bibinfo {pages} {040502} (\bibinfo {year}
  {2010})}\BibitemShut {NoStop}%
\bibitem [{\citenamefont {Oreg}\ \emph {et~al.}(2010)\citenamefont {Oreg},
  \citenamefont {Refael},\ and\ \citenamefont {von Oppen}}]{oreg2010}%
  \BibitemOpen
  \bibfield  {author} {\bibinfo {author} {\bibfnamefont {Y.}~\bibnamefont
  {Oreg}}, \bibinfo {author} {\bibfnamefont {G.}~\bibnamefont {Refael}}, \ and\
  \bibinfo {author} {\bibfnamefont {F.}~\bibnamefont {von Oppen}},\ }\href
  {\doibase 10.1103/PhysRevLett.105.177002} {\bibfield  {journal} {\bibinfo
  {journal} {Phys. Rev. Lett.}\ }\textbf {\bibinfo {volume} {105}},\ \bibinfo
  {pages} {177002} (\bibinfo {year} {2010})}\BibitemShut {NoStop}%
\bibitem [{\citenamefont {Sun}\ and\ \citenamefont {Shah}(2015)}]{sun2015}%
  \BibitemOpen
  \bibfield  {author} {\bibinfo {author} {\bibfnamefont {K.}~\bibnamefont
  {Sun}}\ and\ \bibinfo {author} {\bibfnamefont {N.}~\bibnamefont {Shah}},\
  }\href {\doibase 10.1103/PhysRevB.91.144508} {\bibfield  {journal} {\bibinfo
  {journal} {Phys. Rev. B}\ }\textbf {\bibinfo {volume} {91}},\ \bibinfo
  {pages} {144508} (\bibinfo {year} {2015})}\BibitemShut {NoStop}%
\bibitem [{pri()}]{privatecommunication}%
  \BibitemOpen
  \href@noop {} {}\bibinfo {note} {Francesco Giazotto, Attila Geresdi and
  Charles Marcus, private communications.}\BibitemShut {Stop}%
\bibitem [{\citenamefont {Paajaste}\ \emph {et~al.}(2015)\citenamefont
  {Paajaste}, \citenamefont {Amado}, \citenamefont {Roddaro}, \citenamefont
  {Bergeret}, \citenamefont {Ercolani}, \citenamefont {Sorba},\ and\
  \citenamefont {Giazotto}}]{paajaste2015}%
  \BibitemOpen
  \bibfield  {author} {\bibinfo {author} {\bibfnamefont {J.}~\bibnamefont
  {Paajaste}}, \bibinfo {author} {\bibfnamefont {M.}~\bibnamefont {Amado}},
  \bibinfo {author} {\bibfnamefont {S.}~\bibnamefont {Roddaro}}, \bibinfo
  {author} {\bibfnamefont {F.~S.}\ \bibnamefont {Bergeret}}, \bibinfo {author}
  {\bibfnamefont {D.}~\bibnamefont {Ercolani}}, \bibinfo {author}
  {\bibfnamefont {L.}~\bibnamefont {Sorba}}, \ and\ \bibinfo {author}
  {\bibfnamefont {F.}~\bibnamefont {Giazotto}},\ }\href {\doibase
  10.1021/nl504544s} {\bibfield  {journal} {\bibinfo  {journal} {Nano Lett.}\
  }\textbf {\bibinfo {volume} {15}},\ \bibinfo {pages} {1803} (\bibinfo {year}
  {2015})}\BibitemShut {NoStop}%
\bibitem [{\citenamefont {Giazotto}\ \emph {et~al.}(2011)\citenamefont
  {Giazotto}, \citenamefont {Spathis}, \citenamefont {Roddaro}, \citenamefont
  {Biswas}, \citenamefont {Taddei}, \citenamefont {Governale},\ and\
  \citenamefont {Sorba}}]{giazotto2011}%
  \BibitemOpen
  \bibfield  {author} {\bibinfo {author} {\bibfnamefont {F.}~\bibnamefont
  {Giazotto}}, \bibinfo {author} {\bibfnamefont {P.}~\bibnamefont {Spathis}},
  \bibinfo {author} {\bibfnamefont {S.}~\bibnamefont {Roddaro}}, \bibinfo
  {author} {\bibfnamefont {S.}~\bibnamefont {Biswas}}, \bibinfo {author}
  {\bibfnamefont {F.}~\bibnamefont {Taddei}}, \bibinfo {author} {\bibfnamefont
  {M.}~\bibnamefont {Governale}}, \ and\ \bibinfo {author} {\bibfnamefont
  {L.}~\bibnamefont {Sorba}},\ }\href@noop {} {\bibfield  {journal} {\bibinfo
  {journal} {Nature Phys.}\ }\textbf {\bibinfo {volume} {7}},\ \bibinfo {pages}
  {857} (\bibinfo {year} {2011})}\BibitemShut {NoStop}%
\bibitem [{\citenamefont {Spathis}\ \emph {et~al.}(2011)\citenamefont
  {Spathis}, \citenamefont {Biswas}, \citenamefont {Roddaro}, \citenamefont
  {Sorba}, \citenamefont {Giazotto},\ and\ \citenamefont
  {Beltram}}]{spathis2011}%
  \BibitemOpen
  \bibfield  {author} {\bibinfo {author} {\bibfnamefont {P.}~\bibnamefont
  {Spathis}}, \bibinfo {author} {\bibfnamefont {S.}~\bibnamefont {Biswas}},
  \bibinfo {author} {\bibfnamefont {S.}~\bibnamefont {Roddaro}}, \bibinfo
  {author} {\bibfnamefont {L.}~\bibnamefont {Sorba}}, \bibinfo {author}
  {\bibfnamefont {F.}~\bibnamefont {Giazotto}}, \ and\ \bibinfo {author}
  {\bibfnamefont {F.}~\bibnamefont {Beltram}},\ }\href
  {http://stacks.iop.org/0957-4484/22/i=10/a=105201} {\bibfield  {journal}
  {\bibinfo  {journal} {Nanotechnology}\ }\textbf {\bibinfo {volume} {22}},\
  \bibinfo {pages} {105201} (\bibinfo {year} {2011})}\BibitemShut {NoStop}%
\bibitem [{\citenamefont {Roddaro}\ \emph {et~al.}(2011)\citenamefont
  {Roddaro}, \citenamefont {Pescaglini}, \citenamefont {Ercolani},
  \citenamefont {Sorba}, \citenamefont {Giazotto},\ and\ \citenamefont
  {Beltram}}]{roddaro2011}%
  \BibitemOpen
  \bibfield  {author} {\bibinfo {author} {\bibfnamefont {S.}~\bibnamefont
  {Roddaro}}, \bibinfo {author} {\bibfnamefont {A.}~\bibnamefont {Pescaglini}},
  \bibinfo {author} {\bibfnamefont {D.}~\bibnamefont {Ercolani}}, \bibinfo
  {author} {\bibfnamefont {L.}~\bibnamefont {Sorba}}, \bibinfo {author}
  {\bibfnamefont {F.}~\bibnamefont {Giazotto}}, \ and\ \bibinfo {author}
  {\bibfnamefont {F.}~\bibnamefont {Beltram}},\ }\href@noop {} {\bibfield
  {journal} {\bibinfo  {journal} {Nano Res.}\ }\textbf {\bibinfo {volume}
  {4}},\ \bibinfo {pages} {259} (\bibinfo {year} {2011})}\BibitemShut {NoStop}%
\bibitem [{\citenamefont {van Weperen}\ \emph {et~al.}(2013)\citenamefont {van
  Weperen}, \citenamefont {Plissard}, \citenamefont {Bakkers}, \citenamefont
  {Frolov},\ and\ \citenamefont {Kouwenhoven}}]{vanweperen2013}%
  \BibitemOpen
  \bibfield  {author} {\bibinfo {author} {\bibfnamefont {I.}~\bibnamefont {van
  Weperen}}, \bibinfo {author} {\bibfnamefont {S.~R.}\ \bibnamefont
  {Plissard}}, \bibinfo {author} {\bibfnamefont {E.~P. A.~M.}\ \bibnamefont
  {Bakkers}}, \bibinfo {author} {\bibfnamefont {S.~M.}\ \bibnamefont {Frolov}},
  \ and\ \bibinfo {author} {\bibfnamefont {L.~P.}\ \bibnamefont
  {Kouwenhoven}},\ }\href {\doibase 10.1021/nl3035256} {\bibfield  {journal}
  {\bibinfo  {journal} {Nano Letters}\ }\textbf {\bibinfo {volume} {13}},\
  \bibinfo {pages} {387} (\bibinfo {year} {2013})}\BibitemShut {NoStop}%
\bibitem [{\citenamefont {G\"ul}\ \emph {et~al.}(2015)\citenamefont {G\"ul},
  \citenamefont {van Woerkom}, \citenamefont {van Weperen}, \citenamefont
  {Car}, \citenamefont {Plissard}, \citenamefont {Bakkers},\ and\ \citenamefont
  {Kouwenhoven}}]{gul2015}%
  \BibitemOpen
  \bibfield  {author} {\bibinfo {author} {\bibfnamefont {O.}~\bibnamefont
  {G\"ul}}, \bibinfo {author} {\bibfnamefont {D.~J.}\ \bibnamefont {van
  Woerkom}}, \bibinfo {author} {\bibfnamefont {I.}~\bibnamefont {van Weperen}},
  \bibinfo {author} {\bibfnamefont {D.}~\bibnamefont {Car}}, \bibinfo {author}
  {\bibfnamefont {S.~R.}\ \bibnamefont {Plissard}}, \bibinfo {author}
  {\bibfnamefont {E.~P. A.~M.}\ \bibnamefont {Bakkers}}, \ and\ \bibinfo
  {author} {\bibfnamefont {L.~P.}\ \bibnamefont {Kouwenhoven}},\ }\href
  {http://stacks.iop.org/0957-4484/26/i=21/a=215202} {\bibfield  {journal}
  {\bibinfo  {journal} {Nanotechnology}\ }\textbf {\bibinfo {volume} {26}},\
  \bibinfo {pages} {215202} (\bibinfo {year} {2015})}\BibitemShut {NoStop}%
\bibitem [{\citenamefont {van Weperen}\ \emph {et~al.}(2015)\citenamefont {van
  Weperen}, \citenamefont {Tarasinski}, \citenamefont {Eeltink}, \citenamefont
  {Pribiag}, \citenamefont {Plissard}, \citenamefont {Bakkers}, \citenamefont
  {Kouwenhoven},\ and\ \citenamefont {Wimmer}}]{vanweperen2015}%
  \BibitemOpen
  \bibfield  {author} {\bibinfo {author} {\bibfnamefont {I.}~\bibnamefont {van
  Weperen}}, \bibinfo {author} {\bibfnamefont {B.}~\bibnamefont {Tarasinski}},
  \bibinfo {author} {\bibfnamefont {D.}~\bibnamefont {Eeltink}}, \bibinfo
  {author} {\bibfnamefont {V.~S.}\ \bibnamefont {Pribiag}}, \bibinfo {author}
  {\bibfnamefont {S.~R.}\ \bibnamefont {Plissard}}, \bibinfo {author}
  {\bibfnamefont {E.~P. A.~M.}\ \bibnamefont {Bakkers}}, \bibinfo {author}
  {\bibfnamefont {L.~P.}\ \bibnamefont {Kouwenhoven}}, \ and\ \bibinfo {author}
  {\bibfnamefont {M.}~\bibnamefont {Wimmer}},\ }\href {\doibase
  10.1103/PhysRevB.91.201413} {\bibfield  {journal} {\bibinfo  {journal} {Phys.
  Rev. B}\ }\textbf {\bibinfo {volume} {91}},\ \bibinfo {pages} {201413}
  (\bibinfo {year} {2015})}\BibitemShut {NoStop}%
\bibitem [{\citenamefont {Dubos}\ \emph {et~al.}(2001)\citenamefont {Dubos},
  \citenamefont {Courtois}, \citenamefont {Pannetier}, \citenamefont {Wilhelm},
  \citenamefont {Zaikin},\ and\ \citenamefont {Sch\"on}}]{dubos2001}%
  \BibitemOpen
  \bibfield  {author} {\bibinfo {author} {\bibfnamefont {P.}~\bibnamefont
  {Dubos}}, \bibinfo {author} {\bibfnamefont {H.}~\bibnamefont {Courtois}},
  \bibinfo {author} {\bibfnamefont {B.}~\bibnamefont {Pannetier}}, \bibinfo
  {author} {\bibfnamefont {F.~K.}\ \bibnamefont {Wilhelm}}, \bibinfo {author}
  {\bibfnamefont {A.~D.}\ \bibnamefont {Zaikin}}, \ and\ \bibinfo {author}
  {\bibfnamefont {G.}~\bibnamefont {Sch\"on}},\ }\href {\doibase
  10.1103/PhysRevB.63.064502} {\bibfield  {journal} {\bibinfo  {journal} {Phys.
  Rev. B}\ }\textbf {\bibinfo {volume} {63}},\ \bibinfo {pages} {064502}
  (\bibinfo {year} {2001})}\BibitemShut {NoStop}%
\bibitem [{\citenamefont {le~Sueur}\ \emph {et~al.}(2008)\citenamefont
  {le~Sueur}, \citenamefont {Joyez}, \citenamefont {Pothier}, \citenamefont
  {Urbina},\ and\ \citenamefont {Esteve}}]{lesueur2008}%
  \BibitemOpen
  \bibfield  {author} {\bibinfo {author} {\bibfnamefont {H.}~\bibnamefont
  {le~Sueur}}, \bibinfo {author} {\bibfnamefont {P.}~\bibnamefont {Joyez}},
  \bibinfo {author} {\bibfnamefont {H.}~\bibnamefont {Pothier}}, \bibinfo
  {author} {\bibfnamefont {C.}~\bibnamefont {Urbina}}, \ and\ \bibinfo {author}
  {\bibfnamefont {D.}~\bibnamefont {Esteve}},\ }\href {\doibase
  10.1103/PhysRevLett.100.197002} {\bibfield  {journal} {\bibinfo  {journal}
  {Phys. Rev. Lett.}\ }\textbf {\bibinfo {volume} {100}},\ \bibinfo {pages}
  {197002} (\bibinfo {year} {2008})}\BibitemShut {NoStop}%
\bibitem [{\citenamefont {Bagrets}\ and\ \citenamefont
  {Altland}(2012)}]{bagrets2012}%
  \BibitemOpen
  \bibfield  {author} {\bibinfo {author} {\bibfnamefont {D.}~\bibnamefont
  {Bagrets}}\ and\ \bibinfo {author} {\bibfnamefont {A.}~\bibnamefont
  {Altland}},\ }\href {\doibase 10.1103/PhysRevLett.109.227005} {\bibfield
  {journal} {\bibinfo  {journal} {Phys. Rev. Lett.}\ }\textbf {\bibinfo
  {volume} {109}},\ \bibinfo {pages} {227005} (\bibinfo {year}
  {2012})}\BibitemShut {NoStop}%
\bibitem [{\citenamefont {Pikulin}\ \emph {et~al.}(2012)\citenamefont
  {Pikulin}, \citenamefont {Dahlhaus}, \citenamefont {Wimmer}, \citenamefont
  {Schomerus},\ and\ \citenamefont {Beenakker}}]{pikulin2012}%
  \BibitemOpen
  \bibfield  {author} {\bibinfo {author} {\bibfnamefont {D.~I.}\ \bibnamefont
  {Pikulin}}, \bibinfo {author} {\bibfnamefont {J.~P.}\ \bibnamefont
  {Dahlhaus}}, \bibinfo {author} {\bibfnamefont {M.}~\bibnamefont {Wimmer}},
  \bibinfo {author} {\bibfnamefont {H.}~\bibnamefont {Schomerus}}, \ and\
  \bibinfo {author} {\bibfnamefont {C.~W.~J.}\ \bibnamefont {Beenakker}},\
  }\href {http://stacks.iop.org/1367-2630/14/i=12/a=125011} {\bibfield
  {journal} {\bibinfo  {journal} {New J. Phys.}\ }\textbf {\bibinfo {volume}
  {14}},\ \bibinfo {pages} {125011} (\bibinfo {year} {2012})}\BibitemShut
  {NoStop}%
\bibitem [{\citenamefont {Neven}\ \emph {et~al.}(2013)\citenamefont {Neven},
  \citenamefont {Bagrets},\ and\ \citenamefont {Altland}}]{neven2013}%
  \BibitemOpen
  \bibfield  {author} {\bibinfo {author} {\bibfnamefont {P.}~\bibnamefont
  {Neven}}, \bibinfo {author} {\bibfnamefont {D.}~\bibnamefont {Bagrets}}, \
  and\ \bibinfo {author} {\bibfnamefont {A.}~\bibnamefont {Altland}},\ }\href
  {http://stacks.iop.org/1367-2630/15/i=5/a=055019} {\bibfield  {journal}
  {\bibinfo  {journal} {New J. Phys.}\ }\textbf {\bibinfo {volume} {15}},\
  \bibinfo {pages} {055019} (\bibinfo {year} {2013})}\BibitemShut {NoStop}%
\bibitem [{\citenamefont {Bergeret}\ and\ \citenamefont
  {Tokatly}(2013)}]{PhysRevLett.110.117003}%
  \BibitemOpen
  \bibfield  {author} {\bibinfo {author} {\bibfnamefont {F.~S.}\ \bibnamefont
  {Bergeret}}\ and\ \bibinfo {author} {\bibfnamefont {I.~V.}\ \bibnamefont
  {Tokatly}},\ }\href@noop {} {\bibfield  {journal} {\bibinfo  {journal} {Phys.
  Rev. Lett.}\ }\textbf {\bibinfo {volume} {110}},\ \bibinfo {pages} {117003}
  (\bibinfo {year} {2013})}\BibitemShut {NoStop}%
\bibitem [{\citenamefont {Bergeret}\ and\ \citenamefont
  {Tokatly}(2014)}]{PhysRevB.89.134517}%
  \BibitemOpen
  \bibfield  {author} {\bibinfo {author} {\bibfnamefont {F.~S.}\ \bibnamefont
  {Bergeret}}\ and\ \bibinfo {author} {\bibfnamefont {I.~V.}\ \bibnamefont
  {Tokatly}},\ }\href@noop {} {\bibfield  {journal} {\bibinfo  {journal} {Phys.
  Rev. B}\ }\textbf {\bibinfo {volume} {89}},\ \bibinfo {pages} {134517}
  (\bibinfo {year} {2014})}\BibitemShut {NoStop}%
\bibitem [{\citenamefont {Gorini}\ \emph {et~al.}(2010)\citenamefont {Gorini},
  \citenamefont {Schwab}, \citenamefont {Raimondi},\ and\ \citenamefont
  {Shelankov}}]{gorini10}%
  \BibitemOpen
  \bibfield  {author} {\bibinfo {author} {\bibfnamefont {C.}~\bibnamefont
  {Gorini}}, \bibinfo {author} {\bibfnamefont {P.}~\bibnamefont {Schwab}},
  \bibinfo {author} {\bibfnamefont {R.}~\bibnamefont {Raimondi}}, \ and\
  \bibinfo {author} {\bibfnamefont {A.~L.}\ \bibnamefont {Shelankov}},\ }\href
  {\doibase 10.1103/PhysRevB.82.195316} {\bibfield  {journal} {\bibinfo
  {journal} {Phys. Rev. B}\ }\textbf {\bibinfo {volume} {82}},\ \bibinfo
  {pages} {195316} (\bibinfo {year} {2010})}\BibitemShut {NoStop}%
\bibitem [{\citenamefont {Usadel}(1970)}]{Rev.Lett.25.507}%
  \BibitemOpen
  \bibfield  {author} {\bibinfo {author} {\bibfnamefont {K.~D.}\ \bibnamefont
  {Usadel}},\ }\href@noop {} {\bibfield  {journal} {\bibinfo  {journal} {Phys.
  Rev. Lett.}\ }\textbf {\bibinfo {volume} {25}},\ \bibinfo {pages} {507}
  (\bibinfo {year} {1970})}\BibitemShut {NoStop}%
\bibitem [{\citenamefont {Belzig}\ \emph {et~al.}(1999)\citenamefont {Belzig},
  \citenamefont {Wilhelm}, \citenamefont {Bruder}, \citenamefont {Sch\"on},\
  and\ \citenamefont {Zaikin}}]{SuperlatticesMicrost.25.1251}%
  \BibitemOpen
  \bibfield  {author} {\bibinfo {author} {\bibfnamefont {W.}~\bibnamefont
  {Belzig}}, \bibinfo {author} {\bibfnamefont {F.~K.}\ \bibnamefont {Wilhelm}},
  \bibinfo {author} {\bibfnamefont {C.}~\bibnamefont {Bruder}}, \bibinfo
  {author} {\bibfnamefont {G.}~\bibnamefont {Sch\"on}}, \ and\ \bibinfo
  {author} {\bibfnamefont {A.~D.}\ \bibnamefont {Zaikin}},\ }\href {\doibase
  http://dx.doi.org/10.1006/spmi.1999.0710} {\bibfield  {journal} {\bibinfo
  {journal} {Superlattices Microst.}\ }\textbf {\bibinfo {volume} {25}},\
  \bibinfo {pages} {1251} (\bibinfo {year} {1999})}\BibitemShut {NoStop}%
\bibitem [{rep()}]{representationnote}%
  \BibitemOpen
  \href@noop {} {}\bibinfo {note} {Note that our representation differs by that
  used for example in Refs.~\onlinecite{SuperlatticesMicrost.25.1251} and
  \onlinecite{RevModPhys.58.323}. The two representations are linked via a
  unitary transformation $\hat{U}=(1 - i \hat{\tau}_3 \bar{\sigma}_2)(1 + i
  \bar{\sigma}_2) / 2$ to the Green's functions and the components of the
  Usadel equation.}\BibitemShut {Stop}%
\bibitem [{\citenamefont {Abrikosov}\ and\ \citenamefont
  {Gor'kov}(1962)}]{abrikosov62}%
  \BibitemOpen
  \bibfield  {author} {\bibinfo {author} {\bibfnamefont {A.}~\bibnamefont
  {Abrikosov}}\ and\ \bibinfo {author} {\bibfnamefont {L.}~\bibnamefont
  {Gor'kov}},\ }\href@noop {} {\bibfield  {journal} {\bibinfo  {journal} {Sov.
  Phys. JETP}\ }\textbf {\bibinfo {volume} {15}},\ \bibinfo {pages} {752}
  (\bibinfo {year} {1962})}\BibitemShut {NoStop}%
\bibitem [{\citenamefont {Eschrig}(2000)}]{Phys.Rev.B.61.9061}%
  \BibitemOpen
  \bibfield  {author} {\bibinfo {author} {\bibfnamefont {M.}~\bibnamefont
  {Eschrig}},\ }\href@noop {} {\bibfield  {journal} {\bibinfo  {journal} {Phys.
  Rev. B}\ }\textbf {\bibinfo {volume} {61}},\ \bibinfo {pages} {9061}
  (\bibinfo {year} {2000})}\BibitemShut {NoStop}%
\bibitem [{\citenamefont {Matsubara}(1955)}]{Matsubara}%
  \BibitemOpen
  \bibfield  {author} {\bibinfo {author} {\bibfnamefont {T.}~\bibnamefont
  {Matsubara}},\ }\href {\doibase 10.1143/PTP.14.351} {\bibfield  {journal}
  {\bibinfo  {journal} {Prog. Theor. Phys.}\ }\textbf {\bibinfo {volume}
  {14}},\ \bibinfo {pages} {351} (\bibinfo {year} {1955})}\BibitemShut
  {NoStop}%
\bibitem [{\citenamefont {Cayao}\ \emph {et~al.}(2015)\citenamefont {Cayao},
  \citenamefont {Prada}, \citenamefont {San-Jose},\ and\ \citenamefont
  {Aguado}}]{cayao15}%
  \BibitemOpen
  \bibfield  {author} {\bibinfo {author} {\bibfnamefont {J.}~\bibnamefont
  {Cayao}}, \bibinfo {author} {\bibfnamefont {E.}~\bibnamefont {Prada}},
  \bibinfo {author} {\bibfnamefont {P.}~\bibnamefont {San-Jose}}, \ and\
  \bibinfo {author} {\bibfnamefont {R.}~\bibnamefont {Aguado}},\ }\href
  {\doibase 10.1103/PhysRevB.91.024514} {\bibfield  {journal} {\bibinfo
  {journal} {Phys. Rev. B}\ }\textbf {\bibinfo {volume} {91}},\ \bibinfo
  {pages} {024514} (\bibinfo {year} {2015})}\BibitemShut {NoStop}%
\bibitem [{\citenamefont {Rammer}\ and\ \citenamefont
  {Smith}(1986)}]{RevModPhys.58.323}%
  \BibitemOpen
  \bibfield  {author} {\bibinfo {author} {\bibfnamefont {J.}~\bibnamefont
  {Rammer}}\ and\ \bibinfo {author} {\bibfnamefont {H.}~\bibnamefont {Smith}},\
  }\href@noop {} {\bibfield  {journal} {\bibinfo  {journal} {Rev. Mod. Phys.}\
  }\textbf {\bibinfo {volume} {58}},\ \bibinfo {pages} {323} (\bibinfo {year}
  {1986})}\BibitemShut {NoStop}%
\bibitem [{\citenamefont {Heikkil\"a}\ \emph {et~al.}(2002)\citenamefont
  {Heikkil\"a}, \citenamefont {S\"arkk\"a},\ and\ \citenamefont
  {Wilhelm}}]{Heikkila02}%
  \BibitemOpen
  \bibfield  {author} {\bibinfo {author} {\bibfnamefont {T.~T.}\ \bibnamefont
  {Heikkil\"a}}, \bibinfo {author} {\bibfnamefont {J.}~\bibnamefont
  {S\"arkk\"a}}, \ and\ \bibinfo {author} {\bibfnamefont {F.~K.}\ \bibnamefont
  {Wilhelm}},\ }\href {\doibase 10.1103/PhysRevB.66.184513} {\bibfield
  {journal} {\bibinfo  {journal} {Phys. Rev. B}\ }\textbf {\bibinfo {volume}
  {66}},\ \bibinfo {pages} {184513} (\bibinfo {year} {2002})}\BibitemShut
  {NoStop}%
\bibitem [{\citenamefont {Arjoranta}(2014)}]{juhongradu}%
  \BibitemOpen
  \bibfield  {author} {\bibinfo {author} {\bibfnamefont {J.}~\bibnamefont
  {Arjoranta}},\ }\emph {\bibinfo {title} {Spin-orbit coupling in
  superconductor-normal metal-superconductor junctions}},\ \href@noop {}
  {Master's thesis},\ \bibinfo  {school} {University of Helsinki} (\bibinfo
  {year} {2014})\BibitemShut {NoStop}%
\bibitem [{\citenamefont {Jacobsen}\ and\ \citenamefont
  {Linder}(2015)}]{jacobsen15a}%
  \BibitemOpen
  \bibfield  {author} {\bibinfo {author} {\bibfnamefont {S.~H.}\ \bibnamefont
  {Jacobsen}}\ and\ \bibinfo {author} {\bibfnamefont {J.}~\bibnamefont
  {Linder}},\ }\href {\doibase 10.1103/PhysRevB.92.024501} {\bibfield
  {journal} {\bibinfo  {journal} {Phys. Rev. B}\ }\textbf {\bibinfo {volume}
  {92}},\ \bibinfo {pages} {024501} (\bibinfo {year} {2015})}\BibitemShut
  {NoStop}%
\bibitem [{\citenamefont {Jacobsen}\ \emph {et~al.}(2015)\citenamefont
  {Jacobsen}, \citenamefont {Ouassou},\ and\ \citenamefont
  {Linder}}]{jacobsen15b}%
  \BibitemOpen
  \bibfield  {author} {\bibinfo {author} {\bibfnamefont {S.~H.}\ \bibnamefont
  {Jacobsen}}, \bibinfo {author} {\bibfnamefont {J.~A.}\ \bibnamefont
  {Ouassou}}, \ and\ \bibinfo {author} {\bibfnamefont {J.}~\bibnamefont
  {Linder}},\ }\href {\doibase 10.1103/PhysRevB.92.024510} {\bibfield
  {journal} {\bibinfo  {journal} {Phys. Rev. B}\ }\textbf {\bibinfo {volume}
  {92}},\ \bibinfo {pages} {024510} (\bibinfo {year} {2015})}\BibitemShut
  {NoStop}%
\bibitem [{\citenamefont {Buzdin}\ \emph {et~al.}(1982)\citenamefont {Buzdin},
  \citenamefont {Bulaevskii},\ and\ \citenamefont {Panyukov}}]{buzdin82}%
  \BibitemOpen
  \bibfield  {author} {\bibinfo {author} {\bibfnamefont {A.~I.}\ \bibnamefont
  {Buzdin}}, \bibinfo {author} {\bibfnamefont {L.~N.}\ \bibnamefont
  {Bulaevskii}}, \ and\ \bibinfo {author} {\bibfnamefont {S.~V.}\ \bibnamefont
  {Panyukov}},\ }\href@noop {} {\bibfield  {journal} {\bibinfo  {journal} {JETP
  Lett.}\ }\textbf {\bibinfo {volume} {35}},\ \bibinfo {pages} {178} (\bibinfo
  {year} {1982})}\BibitemShut {NoStop}%
\bibitem [{\citenamefont {Krive}\ \emph {et~al.}(2004)\citenamefont {Krive},
  \citenamefont {Gorelik}, \citenamefont {Shekhter},\ and\ \citenamefont
  {Jonson}}]{krive04}%
  \BibitemOpen
  \bibfield  {author} {\bibinfo {author} {\bibfnamefont {I.~V.}\ \bibnamefont
  {Krive}}, \bibinfo {author} {\bibfnamefont {L.~Y.}\ \bibnamefont {Gorelik}},
  \bibinfo {author} {\bibfnamefont {R.~I.}\ \bibnamefont {Shekhter}}, \ and\
  \bibinfo {author} {\bibfnamefont {M.}~\bibnamefont {Jonson}},\ }\href
  {\doibase http://dx.doi.org/10.1063/1.1739160} {\bibfield  {journal}
  {\bibinfo  {journal} {Low Temp. Phys.}\ }\textbf {\bibinfo {volume} {30}},\
  \bibinfo {pages} {398} (\bibinfo {year} {2004})}\BibitemShut {NoStop}%
\bibitem [{\citenamefont {Bergeret}\ and\ \citenamefont
  {Tokatly}()}]{bergeret14}%
  \BibitemOpen
  \bibfield  {author} {\bibinfo {author} {\bibfnamefont {F.}~\bibnamefont
  {Bergeret}}\ and\ \bibinfo {author} {\bibfnamefont {I.}~\bibnamefont
  {Tokatly}},\ }\href@noop {} {}\bibinfo {note} {[arXiv:1409.4563]}\BibitemShut
  {NoStop}%
\bibitem [{\citenamefont {Zhou}\ \emph {et~al.}(1998)\citenamefont {Zhou},
  \citenamefont {Charlat}, \citenamefont {Spivak},\ and\ \citenamefont
  {Pannetier}}]{zhou98}%
  \BibitemOpen
  \bibfield  {author} {\bibinfo {author} {\bibfnamefont {F.}~\bibnamefont
  {Zhou}}, \bibinfo {author} {\bibfnamefont {P.}~\bibnamefont {Charlat}},
  \bibinfo {author} {\bibfnamefont {B.}~\bibnamefont {Spivak}}, \ and\ \bibinfo
  {author} {\bibfnamefont {B.}~\bibnamefont {Pannetier}},\ }\href {\doibase
  10.1023/A:1022628927203} {\bibfield  {journal} {\bibinfo  {journal} {J. Low
  Temp. Phys.}\ }\textbf {\bibinfo {volume} {110}},\ \bibinfo {pages} {841}
  (\bibinfo {year} {1998})}\BibitemShut {NoStop}%
\bibitem [{sup()}]{supplement}%
  \BibitemOpen
  \href@noop {} {}\bibinfo {note} {See
  http://users.jyu.fi/$\sim$ttheikki/soDOS/ for more examples on the form of
  the local density of states.}\BibitemShut {Stop}%
\bibitem [{\citenamefont {Tanaka}\ and\ \citenamefont
  {Golubov}(2007)}]{tanaka07}%
  \BibitemOpen
  \bibfield  {author} {\bibinfo {author} {\bibfnamefont {Y.}~\bibnamefont
  {Tanaka}}\ and\ \bibinfo {author} {\bibfnamefont {A.~A.}\ \bibnamefont
  {Golubov}},\ }\href {\doibase 10.1103/PhysRevLett.98.037003} {\bibfield
  {journal} {\bibinfo  {journal} {Phys. Rev. Lett.}\ }\textbf {\bibinfo
  {volume} {98}},\ \bibinfo {pages} {037003} (\bibinfo {year}
  {2007})}\BibitemShut {NoStop}%
\bibitem [{\citenamefont {Konstandin}\ \emph {et~al.}(2005)\citenamefont
  {Konstandin}, \citenamefont {Kopu},\ and\ \citenamefont
  {Eschrig}}]{konstandin05}%
  \BibitemOpen
  \bibfield  {author} {\bibinfo {author} {\bibfnamefont {A.}~\bibnamefont
  {Konstandin}}, \bibinfo {author} {\bibfnamefont {J.}~\bibnamefont {Kopu}}, \
  and\ \bibinfo {author} {\bibfnamefont {M.}~\bibnamefont {Eschrig}},\ }\href
  {\doibase 10.1103/PhysRevB.72.140501} {\bibfield  {journal} {\bibinfo
  {journal} {Phys. Rev. B}\ }\textbf {\bibinfo {volume} {72}},\ \bibinfo
  {pages} {140501} (\bibinfo {year} {2005})}\BibitemShut {NoStop}%
\bibitem [{\citenamefont {Alidoust}\ \emph {et~al.}(2010)\citenamefont
  {Alidoust}, \citenamefont {Rashedi}, \citenamefont {Linder},\ and\
  \citenamefont {Sudb\o{}}}]{alidoust10}%
  \BibitemOpen
  \bibfield  {author} {\bibinfo {author} {\bibfnamefont {M.}~\bibnamefont
  {Alidoust}}, \bibinfo {author} {\bibfnamefont {G.}~\bibnamefont {Rashedi}},
  \bibinfo {author} {\bibfnamefont {J.}~\bibnamefont {Linder}}, \ and\ \bibinfo
  {author} {\bibfnamefont {A.}~\bibnamefont {Sudb\o{}}},\ }\href {\doibase
  10.1103/PhysRevB.82.014532} {\bibfield  {journal} {\bibinfo  {journal} {Phys.
  Rev. B}\ }\textbf {\bibinfo {volume} {82}},\ \bibinfo {pages} {014532}
  (\bibinfo {year} {2010})}\BibitemShut {NoStop}%
\bibitem [{\citenamefont {Alidoust}\ \emph {et~al.}(2015)\citenamefont
  {Alidoust}, \citenamefont {Halterman},\ and\ \citenamefont
  {Valls}}]{alidoust15}%
  \BibitemOpen
  \bibfield  {author} {\bibinfo {author} {\bibfnamefont {M.}~\bibnamefont
  {Alidoust}}, \bibinfo {author} {\bibfnamefont {K.}~\bibnamefont {Halterman}},
  \ and\ \bibinfo {author} {\bibfnamefont {O.~T.}\ \bibnamefont {Valls}},\
  }\href {\doibase 10.1103/PhysRevB.92.014508} {\bibfield  {journal} {\bibinfo
  {journal} {Phys. Rev. B}\ }\textbf {\bibinfo {volume} {92}},\ \bibinfo
  {pages} {014508} (\bibinfo {year} {2015})}\BibitemShut {NoStop}%
\bibitem [{\citenamefont {Shelankov}(1985)}]{J.Low.Temp.Phys.60.29}%
  \BibitemOpen
  \bibfield  {author} {\bibinfo {author} {\bibfnamefont {A.}~\bibnamefont
  {Shelankov}},\ }\href@noop {} {\bibfield  {journal} {\bibinfo  {journal} {J.
  Low Temp. Phys.}\ }\textbf {\bibinfo {volume} {60}},\ \bibinfo {pages} {29}
  (\bibinfo {year} {1985})}\BibitemShut {NoStop}%
\end{thebibliography}%

\end{document}